# Evolutionary study of complex organic molecules in high-mass star-forming regions

A. Coletta[1]⋆, F. Fontani[2], V. M. Rivilla[2], C. Mininni[1,2], L. Colzi[1,2], Á. Sánchez-Monge[3], and M. T. Beltrán[2]

[1] Dipartimento di Fisica e Astronomia, Università degli Studi di Firenze, Via Sansone 1, 50019 Sesto Fiorentino (FI), Italy
[2] INAF - Osservatorio Astrofisico di Arcetri, Largo E. Fermi 5, 50125 Firenze, Italy
[3] I. Physikalisches Institut, Universität zu Köln, Zülpicher Str. 77, 50937 Köln, Germany

**ABSTRACT**

We have studied four complex organic molecules (COMs), the oxygen-bearing methyl formate ($CH_3OCHO$) and dimethyl ether ($CH_3OCH_3$) as well as the nitrogen-bearing formamide ($NH_2CHO$) and ethyl cyanide ($C_2H_5CN$), towards a large sample of 39 high-mass star-forming regions representing different evolutionary stages, from early to evolved phases. We aim to identify potential correlations and chemical links between the molecules and to trace their evolutionary sequence through the star formation process.
We analysed spectra obtained at 3, 2, and 0.9 mm with the IRAM-30m telescope. We derived the main physical parameters for each species by fitting the molecular lines. We compared them and evaluated their evolution while also taking several other interstellar environments into account.
We report detections in 20 sources, revealing a clear dust absorption effect on column densities. Derived abundances range between $\sim 10^{-10} - 10^{-7}$ for $CH_3OCHO$ and $CH_3OCH_3$, $\sim 10^{-12} - 10^{-10}$ for $NH_2CHO$, and $\sim 10^{-11} - 10^{-9}$ for $C_2H_5CN$. The abundances of $CH_3OCHO$, $CH_3OCH_3$, and $C_2H_5CN$ are very strongly correlated ($r \geq 0.92$) across $\sim 4$ orders of magnitude. We note that $CH_3OCHO$ and $CH_3OCH_3$ show the strongest correlations in most parameters, and a nearly constant ratio ($\sim 1$) over a remarkable $\sim 9$ orders of magnitude in luminosity for the following wide variety of sources: pre-stellar to evolved cores, low- to high-mass objects, shocks, Galactic clouds, and comets. This indicates that COMs chemistry is likely early developed and then preserved through evolved phases. Moreover, the molecular abundances clearly increase with evolution, covering $\sim 6$ orders of magnitude in the luminosity/mass ratio.
We consider $CH_3OCHO$ and $CH_3OCH_3$ to be most likely chemically linked. They could, for example, share a common precursor, or be formed one from the other. Based on correlations, ratios, and the evolutionary trend, we propose a general scenario for all COMs, involving a formation in the cold, earliest phases of star formation and a following increasing desorption with the progressive thermal and shock-induced heating of the evolving core.

## 1. Introduction

It is well known that most stars are born within crowded clusters (see e.g. Carpenter 2000; Lada & Lada 2003) which also include massive stars ($M_\star \geq 8\,M_\odot$, see e.g. Rivilla et al. 2013, 2014). There is evidence that this could have also been the case for the origin of our Sun (Adams 2010; Pfalzner et al. 2015). Hence, the study of the physical and chemical properties of high-mass star-forming regions can give us important information about the birth environment of our own planetary system.

The formation of high-mass stars takes place in dense and compact cores ($n \geq 10^5$ cm$^{-3}$, $D \leq 0.1$ pc) within interstellar molecular clouds (see e.g. Garay & Lizano 1999; Kurtz et al. 2000; Beuther et al. 2007; Cesaroni et al. 2007; Zinnecker & Yorke 2007; Tan et al. 2014; Yamamoto 2017). The regions involved undergo an evolution following the birth and development of the central star(s), during which their physical (e.g. temperature, density, luminosity) and chemical properties gradually change (see e.g. Caselli 2005; Beuther 2007). Observationally, the first stage can be identified with a high-mass starless core (HMSC), that is, a cold ($T \simeq 15 - 20$ K), dense ($n \simeq 10^5$ cm$^{-3}$), and massive molecular condensation without evidence of star formation activity, which can potentially collapse due to gravitational instability (Tan et al. 2013a,b). The following phase, involving the high-mass protostellar object (HMPO), marks the formation of a protostar within a hot molecular core (HMC), with $T \geq 100$ K and $n \geq 10^7$ cm$^{-3}$ (Kurtz et al. 2000; Fontani et al. 2007; Beltrán et al. 2009). As soon as the (proto)star ignites and starts to heat up the surrounding medium by irradiating its energy, the emitted ultraviolet (UV) photons ionise hydrogen, thus forming an HII region in the close proximity of the star. Driven by the stellar radiation pressure, the ionisation front expands supersonically, so that smaller HII regions are believed to be associated with younger massive stars. Hoare et al. (2007) propose a coarse classification of HII regions based on their size and electron density: The most compact ($R \leq 0.05$ pc) and dense ($n \geq 10^6$ cm$^{-3}$) ones are called hypercompact HII regions (HCHIIs); those with a size of $0.05 < R \leq 0.1$ pc and a density of $n \geq 10^4$ cm$^{-3}$ are called ultracompact HII regions (UCHIIs); finally, those larger than 0.1 pc are compact or classical HII regions (Wood & Churchwell 1989; Kurtz et al. 1994a,b).

An increasing chemical complexity in the molecular environment is frequently observed during this evolution (see e.g. van Dishoeck & Blake 1998; Tan et al. 2014), similarly to what occurs during low-mass star formation (see e.g. Caselli & Ceccarelli 2012; Yamamoto 2017). Hot cores, in particular, show a rich chemistry and the biggest variety of complex organic molecules

⋆ E-mail: `alessandro.coletta@stud.unifi.it`

(COMs, e.g. Caselli 2005; Fontani et al. 2007; Bisschop et al. 2007; Choudhury et al. 2015; Rivilla et al. 2017a,b), which are defined as molecules with six or more atoms including carbon (Herbst & van Dishoeck 2009). High-mass star-forming regions are therefore a very suitable laboratory to study astrochemistry, and particularly the formation of COMs.

COMs are expected to have an important role in prebiotic chemistry, as keys to the formation of basic ingredients of life such as aminoacids, sugars and lipids (Caselli & Ceccarelli 2012; Rivilla et al. 2017a). About 70 different COMs have been identified to date in the interstellar medium (ISM) and circumstellar shells[1]. Since the molecular transitions (at radio-mm as well as FIR and sub-mm wavelengths) are sensitive to the local physical parameters of the gas (temperature and density), the detection of different species allows us to trace zones with different physical conditions within molecular clouds, thus getting considerable information about the formation and destruction pathways of COMs, and the local evolving physics of star-forming regions. However, the mechanisms responsible for the formation of COMs are still under debate. Two main pathways have been proposed: i) gas-phase chemical reactions (see e.g. Duley & Williams 1984; Caselli 2005; Vasyunin & Herbst 2013; Balucani et al. 2015; Skouteris et al. 2018), and ii) surface chemistry on the surface of interstellar dust grains (see e.g. Hasegawa et al. 1992; Ruffle & Herbst 2000; Caselli 2005; Bisschop et al. 2007; Garrod et al. 2008; Ruaud et al. 2015). These processes are not completely independent, but rather complementary, considering how the evolving physical conditions during the star formation affect local chemistry. For example, the grain composition can influence the surrounding gas-phase chemical complexity through desorption. Hence, molecular abundances at different phases of star formation should be strictly related (Garrod & Herbst 2006; Garrod et al. 2008), and a trend with the evolutionary stage of the sources is expected. Investigations about the chemical evolution of star-forming regions at different evolutionary stages have been conducted in several works (e.g. Doty et al. 2002; Beuther et al. 2009; Hoq et al. 2013; Fontani et al. 2011, 2015b; Gerner et al. 2014; Choudhury et al. 2015; Colzi et al. 2018a) using observations of selected simple or complex molecules (and their isotopologues). These works show that the varying physical conditions in the molecular environment during the star-forming process can significantly affect the molecular abundances and their emission lines strength. However, a systematic study of the evolution of COMs within the star formation process in high-mass star-forming regions is still missing.

In this paper we present a study of four COMs in a sample of 39 high-mass star-forming regions representing different evolutionary stages, from HMSCs to UCHIIs. The analysis has two main goals: i) to compare the physical parameters obtained from the emission lines of each molecule (e.g. column density, molecular abundance and excitation temperature), in order to find potential correlations and links between these COMs (such as common pathways or similar physical conditions for their formation); and ii) to evaluate the variation of their abundance with source luminosities and evolutionary stages, in order to find if it can be used as an evolutionary tracer within the star formation process.

In particular, we analyse single-dish observations of the oxygen-bearing molecules $CH_3OCHO$ (methyl formate, hereafter MF) and $CH_3OCH_3$ (dimethyl ether, hereafter DE) (see e.g. Garrod & Herbst 2006; Peeters et al. 2006; Brouillet et al. 2013; Skouteris et al. 2019), and the nitrogen-bearing molecules $NH_2CHO$ (formamide, hereafter F) and $C_2H_5CN$ (ethyl cyanide, hereafter EC) (see e.g. Johnson et al. 1977; Saladino et al. 2012; Adande et al. 2013; López-Sepulcre et al. 2015, 2019; Allen et al. 2018).

Several authors have searched for correlations between the abundances of various O-bearing and N-bearing complex molecules, reporting different, sometimes conflicting results (see e.g. Blake et al. 1987; Caselli et al. 1993; Fontani et al. 2007; Bisschop et al. 2007; Suzuki et al. 2018). In particular, a chemical link between MF and DE is suggested by both recent theoretical models (Garrod & Herbst 2006; Garrod et al. 2008; Garrod 2013; Balucani et al. 2015) and observations (e.g. Bisschop et al. 2007; Brouillet et al. 2013; Jaber et al. 2014; Rivilla et al. 2017a), while a correlation between DE and EC is observed by Fontani et al. (2007) in six HMCs. Bisschop et al. (2007) instead find no correlation between DE and N-bearing species abundances. Moreover, interferometric observations (e.g. Sutton et al. 1995; Blake et al. 1996; Wyrowski et al. 1999; Liu 2005) suggest that O- and N-bearing molecules trace different portions of a molecular star-forming clump (see also Csengeri et al. 2019 and references therein). However, several details are still unclear.

In Section 2 we present our sample, and in Sect. 3 we describe the observations and the data reduction. The molecular line fitting procedure through which we derived the physical parameters is illustrated in Sect. 4. The results are reported in Sect. 5. In Sect. 6 we present an extensive analysis of the results and discuss their potential implications, mainly focusing on correlations among the molecules and with the evolutionary stage of the sources. Lastly, Section 7 summarises the main results of this work and draws the conclusions.

## 2. Source sample

Our sample consists of 39 high-mass star-forming regions, selected to represent different evolutionary stages within the star formation process (from HMSCs to UCHIIs), in order to evaluate the variation of measured molecular parameters through different phases. These sources are part of the sample studied by Fontani et al. (2011, 2014, 2015a,b, 2016, 2018, 2019), Colzi et al. (2018a,b), and Mininni et al. (2018).

The sources for which we have detected at least two molecular transitions of at least one of the COMs studied in this work are 20, and are listed in Table 1. The other 19 sources are listed in Appendix A (Table A.1). We focus the analysis only on the sample of sources with detections. This sample covers a wide range of distances from the Sun ($\sim 1-9$ kpc), luminosities ($\sim 10^3 - 10^7\,L_\odot$) and masses ($\sim 10 - 10^4\,M_\odot$). Sources have been divided into four groups: 1 HMSC, 5 HMPOs, 5 Intermediate (hereafter INTs) and 9 UCHIIs. We based our evolutionary classification on the one made by Fontani et al. (2011) for the sources included in their paper, and on the references listed in Tables 1 and A.1 for the others. HCHII sources 18089-1732 and G75-core, listed as HMPOs in Fontani et al. 2011, have been here included in the INT group. We have defined the INT group to include HCHIIs and high-mass sources in between the HMPO and UCHII phase for which we found uncertain or discordant classifications among different works of literature (see specific references). From an observational point of view, in fact, it is often difficult to clearly differentiate between these kinds of sources, for example due to their structural complexity (see e.g. Beuther et al. 2007).

---

[1] CDMS Catalogue Oct 2019:
https://cdms.astro.uni-koeln.de/classic/molecules

**Table 1.** List of sources with detections of at least one of the COMs studied in this work (MF, DE, F, and EC), sorted by evolutionary stage. Distances from the Sun and bolometric luminosities are taken from literature (see references), while masses have been derived from the molecular hydrogen column densities of the sources from the literature (see Table 4 and Sect. 6.3 for details).

| Source | $\alpha$(J2000) (h : m : s) | $\delta$(J2000) (° : ′ : ″) | $d$ (kpc) | $L$ ($L_\odot$) | $M$ ($M_\odot$) | References |
|---|---|---|---|---|---|---|
| HMSC | | | | | | |
| 05358-mm3 | 05 : 39 : 12.5 | +35 : 45 : 55 | 1.8 | $10^{3.8}$ | $10^{1.9}$ | (1, 6, 31, 32, 33, 34, 35, 36) |
| HMPO | | | | | | |
| AFGL5142-MM | 05 : 30 : 48.0 | +33 : 47 : 54 | 1.8 | $10^{3.6}$ | $10^{1.8}$ | (1, 6, 31, 32, 33, 34, 35, 36) |
| 18182-1433M1 | 18 : 21 : 09.2 | −14 : 31 : 49 | 4.5 | $10^{4.0}$ | $10^{2.5}$ | (2, 3, 4, 34) |
| 18517+0437 | 18 : 54 : 14.2 | +04 : 41 : 41 | 2.9 | $10^{4.1}$ | $10^{2.1}$ | (1, 6, 30, 31, 32, 33, 34, 35, 36) |
| I20293-MM1 | 20 : 31 : 12.8 | +40 : 03 : 23 | 2.0 | $10^{3.6}$ | $10^{1.6}$ | (1, 6, 30, 31, 32, 33, 35) |
| I23385 | 23 : 40 : 54.5 | +61 : 10 : 28 | 4.9 | $10^{4.2}$ | $10^{2.1}$ | (1, 6, 30, 31, 32, 33, 35) |
| INT | | | | | | |
| 18089-1732 | 18 : 11 : 51.4 | −17 : 31 : 28 | 3.6 | $10^{4.5}$ | $10^{2.4}$ | (1, 6, 9, 30, 31, 32, 33, 34, 35, 36) |
| G24.78+0.08 | 18 : 36 : 12.6 | −07 : 12 : 11 | 7.7 | $10^{5.3}$ | $10^{3.5}$ | (2, 10, 11, 34) |
| G31.41+0.31 | 18 : 47 : 34.2 | −01 : 12 : 45 | 3.8 | $10^{4.6}$ | $10^{2.9}$ | (2, 7, 8, 29, 34, 38) |
| 20126+4104M1 | 20 : 14 : 25.9 | +41 : 13 : 34 | 1.7 | $10^{4.0}$ | $10^{2.9}$ | (2, 12, 13, 29, 34, 37, 39) |
| G75-core | 20 : 21 : 44.0 | +37 : 26 : 38 | 3.8 | $10^{4.8}$ | $10^{2.1}$ | (1, 2, 5, 6, 30, 31, 32, 33, 34, 35, 36) |
| UCHII | | | | | | |
| W3(OH) | 02 : 27 : 04.7 | +61 : 52 : 25 | 2.0 | $10^{5.0}$ | $10^{1.5}$ | (16, 20, 21, 22, 34) |
| G5.89-0.39 | 18 : 00 : 30.5 | −24 : 04 : 01 | 1.3 | $10^{5.1}$ | $10^{2.3}$ | (1, 6, 29, 30, 31, 32, 33, 34, 35, 36) |
| G10.47+0.03 | 18 : 08 : 38.0 | −19 : 51 : 50 | 5.8 | $10^{6.1}$ | $10^{3.2}$ | (17, 18, 19, 29, 34) |
| G14.33-0.65 | 18 : 18 : 54.8 | −16 : 47 : 53 | 2.6 | $10^{4.3}$ | $10^{2.5}$ | (2, 23, 24, 34) |
| G29.96-0.02 | 18 : 46 : 03.0 | −02 : 39 : 22 | 8.9 | $10^{5.8}$ | $10^{3.9}$ | (10, 15, 16, 29, 34) |
| G35.20-0.74 | 18 : 58 : 13.0 | +01 : 40 : 36 | 2.2 | $10^{4.5}$ | $10^{2.2}$ | (2, 25, 26, 27, 28, 34) |
| W51 | 19 : 23 : 43.9 | +14 : 30 : 32 | 5.4 | $10^{6.7}$ | $10^{2.8}$ | (7, 14, 34) |
| 19410+2336 | 19 : 43 : 11.4 | +23 : 44 : 06 | 2.1 | $10^{4.0}$ | $10^{2.1}$ | (1, 2, 6, 30, 31, 32, 33, 34, 35, 36) |
| ON1 | 20 : 10 : 09.1 | +31 : 31 : 36 | 2.5 | $10^{4.3}$ | $10^{2.5}$ | (1, 2, 6, 30, 31, 32, 33, 34, 35, 36) |

**References.** [1] Fontani et al. 2011; [2] Colzi et al. 2018b; [3] Rosero et al. 2016; [4] Beuther et al. 2006; [5] Murphy et al. 2010; [6] Fontani et al. 2015a; [7] Rivilla et al. 2017a; [8] De Buizer 2003; [9] Lackington 2011; [10] Cesaroni et al. 2017; [11] Beltrán et al. 2007; [12] Beltrán & de Wit 2016; [13] Cesaroni et al. 1999; [14] Etoka et al. 2012; [15] Kirk et al. 2010; [16] Hoare et al. 2007; [17] Pascucci et al. 2004; [18] López-Sepulcre et al. 2009; [19] Hatchell et al. 2000; [20] Fish & Sjouwerman 2007; [21] Mueller et al. 2002; [22] Wyrowski et al. 1997; [23] Liu et al. 2010; [24] Walsh et al. 1997; [25] Caratti o Garatti et al. 2015; [26] Zhang et al. 2014; [27] Sánchez-Monge et al. 2013; [28] De Buizer 2006; [29] Fontani et al. 2007; [30] Fontani et al. 2014; [31] Fontani et al. 2015b; [32] Fontani et al. 2016; [33] Fontani et al. 2018; [34] Fontani et al. 2019; [35] Colzi et al. 2018a; [36] Mininni et al. 2018; [37] Cesaroni et al. 1997; [38] Immer et al. 2019; [39] Fontani et al. 2006.

## 3. Observations and data reduction

This work uses data obtained with the IRAM-30m[2] Telescope (Pico Veleta, Spain) during three observing sessions: August 2014, June 2015 and December 2016.

We obtained spectra of the 39 high-mass star-forming regions with the EMIR (Eight MIxer Receiver, Carter et al. 2012; Kramer 2016) receiver in bands E090 (E0, at 3 mm), E150 (E1, at 2 mm), and E330 (E3, at 0.9 mm). We investigated nine spectral windows (three at 3 mm, four at 2 mm, and two at 0.9 mm) within the lower side band (LSB) of the receivers. Spectra were obtained with two fast fourier transform spectrometers (FTS, Klein et al. 2012) (see Kramer 1997, 2016): i) FTS200 spectrometer (aggregate bandwidth of 8 GHz), with a 195 kHz frequency resolution, corresponding to $\sim 0.2 - 0.7$ km s$^{-1}$; ii) FTS50 spectrometer (1.8 GHz), with 49 kHz resolution corresponding to $\sim 0.1 - 0.2$ km s$^{-1}$. Table 2 reports the observed frequency ranges (or spectral windows) and the main properties of the setups used for each one. At the observed frequencies, the angular resolution of the telescope (half power beam width), which can be expressed as HPBW(″) = 2460/$\nu$(GHz) (Kramer 2018) is $\sim 27'' - 29''$, $\sim 16'' - 18''$, and $\sim 9''$, for the 3 mm, 2 mm, and 0.9 mm bands, respectively. In more detail, the following setups were employed in each observing run: i) August 2014: E0, E1 receivers and FTS200 spectrometer; ii) June 2015: E0, E1 receivers and FTS50 spectrometer; iii) December 2016: E1, E3 receivers and FTS50 spectrometer. Session i) was performed in position-switching mode, while sessions ii) and iii) were done in wobbler-switching mode, with a maximum wobbler throw of 240″.

The data were reduced using the CLASS software from the GILDAS[3] package (see Pety 2005). First, we converted the measured intensity, originally expressed in antenna temperature units $T_A^*$, into main beam brightness temperature $T_{MB}$, using the relation $T_A^* = T_{MB} \eta_{MB}$, where $\eta_{MB} = B_{\mathrm{eff}}{}^4/F_{\mathrm{eff}}$ is the ratio between the main beam efficiency and the forward efficiency of the telescope. Then, baselines were all removed by fitting the line-free channels with first order polynomial functions, and subtracting them from the spectra.

---

[2] IRAM-30m Documentation: http://www.iram.es/IRAMES/mainWiki/FrontPage

[3] The GILDAS software is available at: http://www.iram.fr/IRAMFR/GILDAS

[4] http://www.iram.es/IRAMES/mainWiki/Iram30mEfficiencies

**Table 2.** List of the spectral windows (first column) observed with the IRAM-30m telescope, with the relative setup used. In order: EMIR receiver with its wavelength, FTS spectrometer with its frequency and velocity resolutions, HPBW of the beam of the telescope at the observed frequencies, and system temperatures for each waveband.

| Frequency (GHz) | Receiver | $\lambda_0$ (mm) | Spectrometer | $\Delta\nu$ (kHz) | $\Delta V_{max}$ (km s$^{-1}$) | HPBW ($''$) | $T_{sys}$ (K) |
|---|---|---|---|---|---|---|---|
| 85.3 – 87.1 | E0 (LSB) | 3 | FTS50 | 49 | 0.2 | | |
| 85.6 – 93.4 | | | FTS200 | 195 | 0.7 | 27 – 29 | 100 – 200 |
| 88.6 – 90.4 | | | FTS50 | 49 | 0.2 | | |
| 140.0 – 141.8 | E1 (LSB) | 2 | FTS50 | 49 | 0.1 | | |
| 141.1 – 148.9 | | | FTS200 | 195 | 0.4 | 16 – 18 | 100 – 500 |
| 143.3 – 145.1 | | | FTS50 | 49 | 0.1 | | |
| 151.8 – 153.6 | | | FTS50 | 49 | 0.1 | | |
| 280.9 – 282.7 | E3 (LSB) | 0.9 | FTS50 | 49 | 0.05 | 9 | 400 – 1000 |
| 284.2 – 286.0 | | | FTS50 | 49 | 0.05 | | |

## 4. Data analysis: Molecular line fitting

Baseline subtracted spectra of the sources were exported from CLASS to MADCUBA[5] (MAdrid Data CUBe Analysis, Martín et al. 2019) to perform the molecular line fitting procedure, in order to estimate the physical parameters of MF, DE, F, and EC. We identified the transitions of each molecule using the SLIM (Spectral Line Identification and LTE Modelling) tool of MADCUBA, which searches the JPL[6] (Pickett et al. 1998) and CDMS[7] (Müller et al. 2005) catalogues for all rotational transitions of the molecules within the spectral windows covered by the data. In particular, the JPL catalogue was used for MF lines, while CDMS for DE, F, and EC lines (Ilyushin et al. 2009; Endres et al. 2009; Kryvda et al. 2009; Brauer et al. 2009 and refs. therein, respectively). Molecules were considered as clearly detected if we could identify at least two of their transitions with peak intensity $T_{MB} \geq 3\sigma$ (where $\sigma$ is the rms noise of the spectrum). SLIM generates a synthetic spectrum of the source, based on the assumption of local thermodynamic equilibrium (LTE) conditions. The LTE assumption is a reasonably good approximation for these star-forming regions, since at their high typical densities ($n > 10^5$ cm$^{-3}$) the molecular energy levels populations are thermalised. The synthetic spectrum considers five input physical parameters: total molecular column density ($N$), excitation temperature ($T_{ex}$), radial systemic velocity of the source with respect to the local standard of rest ($V_{LSR}$), full-width half-maximum (FWHM) of the lines, and angular size of the emission ($\theta$). SLIM assumes that all the transitions of a certain species have the same $V_{LSR}$, FWHM, and $T_{ex}$. By varying the values of the parameters we can model the theoretical profile of the spectrum until the best fit to the observed one is found. The AUTOFIT function of SLIM automatically compares the two spectra, performing a non-linear least-squares fitting via the Levenberg-Marquardt algorithm (see Press et al. 2007; Martín et al. 2019; Rivilla et al. 2019) to provide the optimal combination of the above-mentioned parameters with the associated uncertainties. Other quantities like integrated intensity and opacity ($\tau$) are also computed for each detected transition. In our case, all transitions proved to be optically thin ($\tau \ll 1$).

In the molecular line fitting procedure, a fit including all the three wavebands was first attempted. However, it was not possible to properly fit all the lines simultaneously, due to the differential attenuation caused by dust absorption at each wavelength. This effect is further discussed in Sect. 6.1. Therefore, for each source and molecule, we fitted the three observed wavebands (3, 2, and 0.9 mm) separately. The input parameters have been left free when possible. In some cases, leaving all five parameters free did not allow the algorithm to converge. Hence, we fixed one or more among velocity, FWHM, $\theta$ (and, if necessary, $T_{ex}$) to the values that best reproduced the observed spectrum. As initial guesses for the parameters, we used $T_{ex} = 100$ K, the $V_{LSR}$ known from the observations, and approximate estimates of the source sizes ($\theta_0$, ranging from ~0.6$''$ to ~3.8$''$) derived from their distance ($d$) assuming a diameter $D_0 \simeq 5000$ au.

We have applied the beam dilution factor taking into account $\theta$ and the frequency-dependent beam size (HPBW), thus obtaining source-averaged molecular column densities. To derive the source size for each molecule, we have left free $\theta$ and run AUTOFIT in the 2 mm band, being the frequency range in which we report the most numerous detections (see Sect. 5.1). We have then used the obtained values to perform the fits at 3 mm and 0.9 mm. The source size $\theta$ was left free to vary between molecules, as they might trace different regions within the same source.

A selection of the fits of the detected molecular lines performed in different wavebands and sources is shown in Figs. B.1-B.4 of Appendix B.

## 5. Results

In this Section we present the results obtained for the sources, listed in Table 1, which showed enough transitions to derive the physical parameters for MF, DE, F, and EC. Additional results (see below) are available in Appendix C.

### 5.1. Detection summary

We have detected at least one of the four molecules in 20 of the 39 sources. DE was found in 19 sources, MF in 13, EC in 9, and F in 5 sources. DE has been detected at all stages (1 HMSC, 5 HMPOs, 4 INTs, and 9 UCHIIs), while MF in 2 HMPOs, 4 INTs, and 7 UCHIIs, EC in 3 INTs and 6 UCHIIs, and F in 4 INTs and 1 UCHII. In general, the highest number of detections has been reported in UCHII regions (45%, 9 sources), followed by HMPOs and INTs (25% each, 5 sources), and HMSCs (5%, 1 source). This result is not affected by a distance-induced observational bias, as the average distances of the sources of the different evolutionary groups are consistent (see Table 1). A possible interpretation of the distribution of detections among the different groups is discussed in Sect. 6.4. Table 3 shows, for each observed band, the number of sources in which the molecules were identified. The detected rotational transitions considering all the sources are listed in Tables D.1 (2 mm band), D.2 (0.9 mm), and D.3 (3 mm) of Appendix D. The highest number of detected transitions for all COMs has been reported in the 2 mm band. Being also less affected by dust absorption (see Sect. 6.1) than the 0.9 mm band (the band with the second-highest number of detections), we considered the 2 mm data to be the most reliable, and decided to take them as a reference for our analysis, for instance for the derivation of molecular abundances, as we see in Sect. 5.5.

---
[5] MADCUBA is a software developed in the Madrid Center of Astrobiology (INTA-CSIC) which enables to visualise and analyse single spectra and data cubes; MADCUBA is available at: http://cab.inta-csic.es/madcuba/MADCUBA_IMAGEJ/ImageJMadcuba.html
[6] Jet Propulsion Laboratory catalogue: http://spec.jpl.nasa.gov/
[7] Cologne Database for Molecular Spectroscopy: https://cdms.astro.uni-koeln.de/cdms/portal

**Table 3.** Number of sources with detections per band per molecule, within the sample of 39 sources.

| $\lambda_0$ | # of sources | | | |
|---|---|---|---|---|
| (mm) | MF | DE | F | EC |
| 0.9 | 7 | 14 | 3 | 6 |
| 2 | 13 | 17 | 4 | 9 |
| 3 | 4 | 3 | 2 | 3 |
| TOT. | 13 | 19 | 5 | 9 |

### 5.2. Molecular source sizes

Columns 2-5 of Table 4 show the source angular sizes obtained for each molecule from the fitting procedure. Reported values were derived at 2 mm (as explained in Sect. 4), except for a few cases (see caption of Table 4) in which the molecule was detected only at 0.9 mm. Derived $\theta$ values are consistent with direct high-resolution measurements (interferometric maps) such as the ones presented by Zhang et al. (2002) for AFGL5142-MM ($\theta \simeq 1-2''$), Rivilla et al. (2017a) for G31.41+0.31 ($\sim 1-2''$), and Olmi et al. (2003) and Beltrán et al. (2011) for G29.96-0.02 ($\sim 2''$). In addition, Col. 7 of Table 4 reports the overall ranges of the corresponding linear size (diameter $D$) among the molecules detected in each source. We note that the sizes obtained from different molecules in the same source are similar, and in only two cases they differ by a factor $\sim 2$ at most. This ensures that the derived molecular column densities can be consistently compared. Moreover, the average linear sizes (5800–7300 au) obtained from each molecule considering all the sources are consistent. A more detailed discussion of the molecular source sizes is addressed in Sect. 6.2.3.

### 5.3. Excitation temperatures, FWHM, and systemic velocities

Excitation temperatures ($T_{ex}$) obtained for each molecule assuming LTE conditions (see Sect. 4) are shown in Table C.1 of Appendix C.1. Best-fit values in the three observed wavebands are called $T_1$ (0.9 mm band), $T_2$ (2 mm), and $T_3$ (3 mm). A high variability between the three bands can be noticed in all molecules. Although one could expect on average higher excitation temperatures at higher frequencies (i.e. $T_1 > T_2 > T_3$) because of the higher average energy of the detected transitions (see e.g. Tables D.1-D.3), our results do not show any clear trend with frequency. The small number of transitions detected at 0.9 and 3 mm, particularly compared to the 2 mm band, could have prevented a more accurate determination of $T_{ex}$ in those bands. For this reason, we decided to take the more reliable $T_2$ as reference values for our sources, rerunning the fits at 0.9 and 3 mm with $T_{ex}$ fixed to $T_2$. The values of $T_2$ cover a wide range: $\sim 20-220$ K for MF, $\sim 30-170$ K for DE, $\sim 90-115$ K for F, and $\sim 30-200$ K for EC. Further considerations on excitation temperatures are made in Sect. 6.2.3.

The FWHM of the lines (assumed to be unique for each species in a given source, even when multiple transitions are detected) obtained from the 2 mm fits is listed for each molecule in Table 5. Potential correlations between the FWHM of the molecules are discussed in Sect. 6.2.3.

The other physical parameter derived from the molecular line fitting, the LSR source velocity ($V_{LSR}$), is listed for the 2 mm waveband in Table C.2 of Appendix C.2. The derived values are consistent with what found by other authors, such as Rivilla et al. (2017a) for G31.41+0.31 ($V_{LSR} \simeq 96-98$ km s$^{-1}$), Olmi et al. (2003) for G29.96-0.02 (98 km s$^{-1}$), and Fontani et al. (2019, Table 1, and refs. therein) for the other sources.

### 5.4. Molecular column densities

Source-averaged total column densities ($N$) measured for each detected molecule are given in Tables 6-9. As done for temperatures, they are listed as $N_1$ (0.9 mm band), $N_2$ (2 mm), and $N_3$ (3 mm). We assumed the same source size and excitation temperature (those obtained at 2 mm, see Sects. 5.2 and 5.3, respectively) for all wavebands. This made $N_1$ and $N_3$ more consistent with $N_2$, also reducing their uncertainties (Cols. 2 and 4 of Tables 6-9). Measured column densities range from $\sim 10^{15}$ to $\sim 10^{18}$ cm$^{-2}$ for MF, DE, and EC, with G31.41+0.31 (INT), G10.47+0.03, and W51 (UCHIIs) reporting the highest values, and from $\sim 10^{14}$ to $10^{17}$ cm$^{-2}$ for F, with G31.41+0.31 showing the highest value. For all molecules and sources we observe that $N_3 > N_2 > N_1$. This trend is discussed and explained in Sect. 6.1 and Appendix E, where comparisons between the column densities measured at different wavebands are made.

### 5.5. Molecular abundances

Molecular abundances with respect to molecular hydrogen (H$_2$) have been derived from the total column densities $N_2$ (Col. 3 of Tables 6-9). Molecular hydrogen column densities ($N(H_2)$) and their corresponding angular sizes ($\theta_{H_2}$) were taken from literature (see references in Table 4). All these values are beam-averaged ($\theta_{H_2} \simeq 20''-60''$). Therefore, our source-averaged molecular column densities ($\theta \simeq 1''-3''$, see Table 4) have been rescaled to the respective $\theta_{H_2}$ by multiplying them by the factor $(\theta/\theta_{H_2})^2$:

$$N' = N_2 (\theta/\theta_{H_2})^2. \qquad (1)$$

By doing this, we balanced potential discrepancies between column densities corresponding to different angular scales. Then we computed the fractional abundances of the molecules ($X$) through the formula $X = N'/N(H_2)$. The parameters used to rescale the column densities and obtain the abundances are listed in Table 4 for the 20 sources with detections. Sources 05358-mm3 (HMSC) and 19410+2336 (UCHII) are included for completeness, although for these sources the only species detected, DE, has only been detected at 0.9 mm (see Table 7), so they were not considered for the abundance calculation.

Molecular abundances derived for the 18 sources with detections in the 2 mm band are listed in Table 10. We obtained molecular abundances for 5 HMPOs, 5 INTs, and 8 UCHIIs. In the error estimates we included the molecular column density uncertainties from AUTOFIT, and assumed a reasonable 20% error on the $N(H_2)$ values from literature. Derived fractional abundances range from $\sim 10^{-10}$ to $\sim 10^{-7}$ for MF and DE, from $\sim 10^{-12}$ to $\sim 10^{-10}$ for F, and from $\sim 10^{-11}$ to $\sim 10^{-9}$ for EC. G10.47+0.03 and W51 (UCHII regions) show the highest abundances of MF, DE, and EC, whereas F is most abundant in G31.41+0.31 ($X \simeq 10^{-10}$). The abundances of MF, DE, and EC are consistent with the ones recently predicted for hot cores by Bonfand et al. (2019) through chemical models. DE abundances are also comparable to those observed by Fontani et al. (2007) in several high-mass star-forming regions. MF and EC abundances are consistent with those found by Allen et al. (2018) in G35.20-0.74, while the abundances of F agree with the ones found by Kahane et al. (2013) and López-Sepulcre et al. (2015) in several low- to high-mass star-forming regions.

**Table 4.** Best-fit values of the source angular size for each molecule ($\theta$, obtained at 2 mm except when differently specified[a], see Sect. 4), and average value $\bar{\theta}$. Values without error come from fits performed with the $\theta$ parameter fixed. We also report the overall range of source linear sizes ($D$) corresponding to $\theta$, and for each molecule the average $D$ considering all the sources. For each source, molecular hydrogen column densities ($N(H_2)$) with their angular scale ($\theta_{H_2}$) and reference are also listed. Together with $\theta$, these parameters have been used to rescale the source-averaged molecular column densities and derive the abundances (see Sect. 5.5 and Eq. 1). Here and in the following tables, the horizontal black lines subdivide the sources according to their evolutionary classification (see Table 1).

| Source | $\theta$ (″) MF | $\theta$ (″) DE | $\theta$ (″) F | $\theta$ (″) EC | $\bar{\theta}$ (″) | $D$ ($10^3$ au) | $N(H_2)$ (cm$^{-2}$) | $\theta_{H_2}$ (″) | Ref. |
|---|---|---|---|---|---|---|---|---|---|
| 05358-mm3 |  | 2.4[a] |  |  | 2.4 | 4.3 | $1.1 \cdot 10^{23}$ | 28 | (1) |
| AFGL5142-MM | 2.4 | 2.4 |  |  | 2.4 | 4.3 | $1.0 \cdot 10^{23}$ | 28 | (1) |
| 18182-1433M1 |  | 2.2 |  |  | 2.2 | 9.9 | $3.9 \cdot 10^{22}$ | 36.6 | (5) |
| 18517+0437 | 1.2 | 1.6 |  |  | 1.4 | 3.5 – 4.6 | $7.9 \cdot 10^{22}$ | 28 | (1) |
| I20293-MM1 |  | 2.4 |  |  | 2.4 | 4.8 | $4.9 \cdot 10^{22}$ | 28 | (1) |
| I23385 |  | 1.4 |  |  | 1.4 | 6.9 | $2.4 \cdot 10^{22}$ | 28 | (1) |
| 18089-1732 | $1.1 \pm 0.3$ | $1.2 \pm 0.1$ | 1.2 | 1.4 | 1.2 | 4.0 – 5.0 | $9.6 \cdot 10^{22}$ | 28 | (1) |
| G24.78+0.08 | 1.9 | 1.6 | 1.2 | $1.1 \pm 0.2$ | 1.5 | 8.5 – 14.6 | $1.4 \cdot 10^{23}$ | 36.6 | (5) |
| G31.41+0.31 | $1.3 \pm 0.1$ | $1.7 \pm 0.1$ | 1.3 | $0.9 \pm 0.1$ | 1.3 | 3.4 – 6.5 | $1.4 \cdot 10^{23}$ | 36.6 | (5) |
| 20126+4104M1 |  |  | 2.8 |  | 2.8 | 4.8 | $2.8 \cdot 10^{24}$ | 18 | (4) |
| G75-core | 1.2 | 1.0 |  |  | 1.1 | 3.8 – 4.6 | $4.4 \cdot 10^{22}$ | 28 | (1) |
| W3(OH) | 2.5 | 2.5 |  | 2.4 | 2.5 | 4.8 – 5.0 | $0.5 \cdot 10^{23}$ | 23 | (3) |
| G5.89-0.39 |  | $0.9 \pm 0.2$ |  | 1.2 | 1.1 | 1.2 – 1.6 | $5.5 \cdot 10^{23}$ | 28 | (1) |
| G10.47+0.03 | 1.6 | 1.6 |  | 2.4 | 1.9 | 9.3 – 13.9 | $5.2 \cdot 10^{22}$ | 59 | (2) |
| G14.33-0.65 | 2.0 | 2.0 |  |  | 2.0 | 5.2 | $1.3 \cdot 10^{23}$ | 36.6 | (5) |
| G29.96-0.02 | 1.7 | 1.4 |  | 1.6 | 1.6 | 12.5 – 15.1 | $9.5 \cdot 10^{22}$ | 59 | (2) |
| G35.20-0.74 | 2.4 | 2.4 |  |  | 2.4 | 5.3 | $9.8 \cdot 10^{22}$ | 36.6 | (5) |
| W51 | $1.7 \pm 0.1$ | $1.7 \pm 0.1$ | $1.1^{[a]}$ | $1.7 \pm 0.1$ | 1.6 | 5.9 – 9.2 | $2.0 \cdot 10^{23}$ | 19 | (3) |
| 19410+2336 |  | $1.7^{[a]}$ |  |  | 1.7 | 3.6 | $1.4 \cdot 10^{23}$ | 28 | (1) |
| ON1 | 0.8 | 0.8 |  |  | 0.8 | 2.0 | $5.2 \cdot 10^{22}$ | 59 | (2) |
| average $D$ ($10^3$ au) | $6.7 \pm 1.1$ | $6.0 \pm 0.7$ | $5.8 \pm 0.8$ | $7.3 \pm 1.4$ |  |  |  |  |  |

[a] Source size $\theta$ obtained from the fit at 0.9 mm.
**References.** [1] Fontani et al. 2018; [2] Liu et al. 2010; [3] Rivilla et al. 2016; [4] Fontani et al. 2006; [5] Mininni et al. in prep.

**Table 5.** The FWHM of the lines obtained for each molecule in the 2 mm waveband. Values without error come from fits performed with the FWHM parameter fixed.

| Source | FWHM (km s$^{-1}$) MF | DE | F | EC |
|---|---|---|---|---|
| AFGL5142-MM | $4.8 \pm 0.2$ | $4.6 \pm 0.3$ |  |  |
| 18182-1433M1 |  | 3.0 |  |  |
| 18517+0437 | $6.1 \pm 0.9$ | 3.0 |  |  |
| I20293-MM1 |  | $5.4 \pm 0.7$ |  |  |
| I23385 |  | 3.0 |  |  |
| 18089-1732 | $4.1 \pm 0.1$ | $3.7 \pm 0.1$ | $5.2 \pm 0.7$ | $5.2 \pm 0.2$ |
| G24.78+0.08 | 5.0 | $5.1 \pm 0.2$ | $5.0 \pm 0.6$ | 5.7 |
| G31.41+0.31 | $5.2 \pm 0.1$ | $4.8 \pm 0.1$ | $9.5 \pm 0.3$ | $7.2 \pm 0.2$ |
| 20126+4104M1 |  |  | $8.4 \pm 1.2$ |  |
| G75-core | $3.6 \pm 0.3$ | $3.9 \pm 0.4$ |  |  |
| W3(OH) | $9.6 \pm 0.1$ | $7.9 \pm 0.5$ |  | $7.7 \pm 0.3$ |
| G5.89-0.39 |  | $4.1 \pm 0.4$ |  | $8.7 \pm 0.9$ |
| G10.47+0.03 | $9.3 \pm 0.3$ | $9.4 \pm 0.8$ |  | $10.2 \pm 0.2$ |
| G14.33-0.65 | 1.9 | $4.0 \pm 0.3$ |  | 2.6 |
| G29.96-0.02 | 4.6 | $4.6 \pm 0.3$ |  | $6.5 \pm 0.2$ |
| G35.20-0.74 | 3.7 | 3.0 |  |  |
| W51 | 6.0 | $6.2 \pm 0.1$ |  | $10.8 \pm 0.1$ |
| ON1 | 4.0 | $6.1 \pm 0.8$ |  |  |

Correlations between the molecular abundances of the different COMs, and their behaviour during different evolutionary stages, are investigated and discussed in Sects. 6.2.1 and 6.3, respectively.

**Table 6.** Source-averaged total column densities of MF obtained from the fits (see Sects. 4 and 5.4) in the three observed wavebands: $N_1$ (0.9 mm), $N_2$ (2 mm), and $N_3$ (3 mm).

| Source | $N$(MF) (cm$^{-2}$) $N_1$ | $N_2$ | $N_3$ |
|---|---|---|---|
| AFGL5142-MM |  | $(4.1 \pm 0.8) 10^{15}$ |  |
| 18517+0437 | $(1.3 \pm 0.1) 10^{16}$ | $(4.4 \pm 1.1) 10^{16}$ |  |
| 18089-1732 | $(3.8 \pm 0.5) 10^{16}$ | $(2.5 \pm 0.9) 10^{17}$ | $(6.4 \pm 0.8) 10^{17}$ |
| G24.78+0.08 | $(1.2 \pm 0.1) 10^{16}$ | $(1.7 \pm 0.2) 10^{17}$ | $(4.8 \pm 0.4) 10^{17}$ |
| G31.41+0.31 | $(1.0 \pm 0.1) 10^{17}$ | $(1.8 \pm 0.2) 10^{18}$ | $(3.9 \pm 0.1) 10^{18}$ |
| G75-core |  | $(1.0 \pm 0.1) 10^{16}$ |  |
| W3(OH) | $(2.4 \pm 0.3) 10^{16}$ | $(1.0 \pm 0.1) 10^{17}$ |  |
| G10.47+0.03 | $(1.9 \pm 0.1) 10^{17}$ | $(1.7 \pm 0.1) 10^{18}$ |  |
| G14.33-0.65 |  | $(1.9 \pm 1.5) 10^{16}$ | $(6.6 \pm 1.1) 10^{16}$ |
| G29.96-0.02 |  | $(3.7 \pm 0.9) 10^{16}$ |  |
| G35.20-0.74 |  | $(2.6 \pm 0.6) 10^{16}$ |  |
| W51 | $(4.9 \pm 0.1) 10^{17}$ | $(2.7 \pm 0.1) 10^{18}$ |  |
| ON1 |  | $(3.2 \pm 1.7) 10^{16}$ |  |

**Table 7.** Same as Table 6, but for DE.

| Source | $N$(DE) (cm$^{-2}$) | | |
|---|---|---|---|
| | $N_1$ | $N_2$ | $N_3$ |
| 05358-mm3 | $(1.0 \pm 0.8)\,10^{15}$ | | |
| AFGL5142-MM | $(3.2 \pm 0.1)\,10^{15}$ | $(8.1 \pm 1.2)\,10^{15}$ | |
| 18182-1433M1 | | $(1.9 \pm 0.5)\,10^{16}$ | |
| 18517+0437 | $(1.0 \pm 0.1)\,10^{16}$ | $(1.6 \pm 0.4)\,10^{16}$ | |
| I20293-MM1 | | $(7.1 \pm 1.4)\,10^{15}$ | |
| I23385 | | $(6 \pm 5)\,10^{15}$ | |
| 18089-1732 | $(6.0 \pm 0.1)\,10^{16}$ | $(2.9 \pm 0.3)\,10^{17}$ | $(5.9 \pm 0.5)\,10^{17}$ |
| G24.78+0.08 | $(1.6 \pm 0.1)\,10^{16}$ | $(1.8 \pm 0.2)\,10^{17}$ | $(3.7 \pm 0.3)\,10^{17}$ |
| G31.41+0.31 | $(5.9 \pm 0.2)\,10^{16}$ | $(8.1 \pm 0.6)\,10^{17}$ | $(1.5 \pm 0.1)\,10^{18}$ |
| G75-core | $(1.2 \pm 0.1)\,10^{16}$ | $(2.3 \pm 0.5)\,10^{16}$ | |
| W3(OH) | $(1.9 \pm 0.1)\,10^{16}$ | $(6.8 \pm 0.7)\,10^{16}$ | |
| G5.89-0.39 | $(3.2 \pm 0.3)\,10^{16}$ | $(1.3 \pm 0.4)\,10^{17}$ | |
| G10.47+0.03 | $(2.6 \pm 0.1)\,10^{17}$ | $(1.8 \pm 0.2)\,10^{18}$ | |
| G14.33-0.65 | | $(7.4 \pm 0.8)\,10^{16}$ | |
| G29.96-0.02 | $(1.4 \pm 0.2)\,10^{16}$ | $(2.0 \pm 0.6)\,10^{17}$ | |
| G35.20-0.74 | | $(1.8 \pm 0.4)\,10^{16}$ | |
| W51 | $(6.2 \pm 0.1)\,10^{17}$ | $(1.8 \pm 0.1)\,10^{18}$ | |
| 19410+2336 | $(3.1 \pm 0.9)\,10^{15}$ | | |
| ON1 | $(3.8 \pm 0.2)\,10^{16}$ | $(8.1 \pm 1.5)\,10^{16}$ | |

**Table 8.** Same as Tables 6 and 7, but for F.

| Source | $N$(F) (cm$^{-2}$) | | |
|---|---|---|---|
| | $N_1$ | $N_2$ | $N_3$ |
| 18089-1732 | $(5.0 \pm 0.8)\,10^{14}$ | $(2.6 \pm 1.1)\,10^{15}$ | |
| G24.78+0.08 | | $(4.1 \pm 0.4)\,10^{15}$ | $(4.0 \pm 0.7)\,10^{16}$ |
| G31.41+0.31 | $(9.5 \pm 1.2)\,10^{14}$ | $(3.5 \pm 0.4)\,10^{16}$ | $(1.0 \pm 0.1)\,10^{17}$ |
| 20126+4104M1 | | $(7 \pm 3)\,10^{14}$ | |
| W51 | $(3.3 \pm 0.3)\,10^{16}$ | | |

**Table 9.** Same as Tables 6-8, but for EC.

| Source | $N$(EC) (cm$^{-2}$) | | |
|---|---|---|---|
| | $N_1$ | $N_2$ | $N_3$ |
| 18089-1732 | $(2.3 \pm 0.1)\,10^{15}$ | $(8.6 \pm 0.3)\,10^{15}$ | $(2.3 \pm 0.1)\,10^{16}$ |
| G24.78+0.08 | $(4.4 \pm 0.4)\,10^{15}$ | $(2.0 \pm 1.0)\,10^{16}$ | $(4.0 \pm 0.3)\,10^{16}$ |
| G31.41+0.31 | $(1.4 \pm 0.1)\,10^{16}$ | $(3.3 \pm 0.5)\,10^{17}$ | $(1.0 \pm 0.1)\,10^{18}$ |
| W3(OH) | $(1.1 \pm 0.1)\,10^{15}$ | $(4.1 \pm 0.4)\,10^{15}$ | |
| G5.89-0.39 | | $(7.9 \pm 1.7)\,10^{15}$ | |
| G10.47+0.03 | $(5.4 \pm 0.1)\,10^{15}$ | $(4.6 \pm 0.1)\,10^{16}$ | |
| G14.33-0.65 | | $(8 \pm 3)\,10^{14}$ | |
| G29.96-0.02 | | $(5.9 \pm 0.3)\,10^{15}$ | |
| W51 | $(1.1 \pm 0.1)\,10^{16}$ | $(7.3 \pm 0.5)\,10^{16}$ | |

**Table 10.** Abundances with respect to H$_2$ of MF, DE, F, and EC, derived (see Sect. 5.5) from the total column densities in the 2 mm band (Tables 6-9).

| Source | $X = N'/N(H_2)$ | | | |
|---|---|---|---|---|
| | MF | DE | F | EC |
| AFGL5142-MM | $(3.0 \pm 1.2)\,10^{-10}$ | $(6 \pm 2)\,10^{-10}$ | | |
| 18182-1433M1 | | $(1.7 \pm 0.8)\,10^{-9}$ | | |
| 18517+0437 | $(1.0 \pm 0.5)\,10^{-9}$ | $(7 \pm 3)\,10^{-10}$ | | |
| I20293-MM1 | | $(1.1 \pm 0.4)\,10^{-9}$ | | |
| I23385 | | $(6 \pm 6)\,10^{-10}$ | | |
| 18089-1732 | $(4 \pm 2)\,10^{-9}$ | $(5.5 \pm 1.7)\,10^{-9}$ | $(5 \pm 3)\,10^{-11}$ | $(2.2 \pm 0.5)\,10^{-10}$ |
| G24.78+0.08 | $(3.3 \pm 1.1)\,10^{-9}$ | $(2.4 \pm 0.7)\,10^{-9}$ | $(3.1 \pm 0.9)\,10^{-11}$ | $(1.3 \pm 0.9)\,10^{-10}$ |
| G31.41+0.31 | $(1.7 \pm 0.5)\,10^{-8}$ | $(1.3 \pm 0.3)\,10^{-8}$ | $(3.2 \pm 1.0)\,10^{-10}$ | $(1.4 \pm 0.5)\,10^{-9}$ |
| 20126+4104M1 | | | $(6 \pm 4)\,10^{-12}$ | |
| G75-core | $(4.4 \pm 1.1)\,10^{-10}$ | $(7 \pm 3)\,10^{-10}$ | | |
| W3(OH) | $(2.4 \pm 0.6)\,10^{-8}$ | $(1.6 \pm 0.5)\,10^{-8}$ | | $(9 \pm 3)\,10^{-10}$ |
| G5.89-0.39 | | $(2.5 \pm 1.2)\,10^{-10}$ | | $(2.6 \pm 1.1)\,10^{-11}$ |
| G10.47+0.03 | $(2.5 \pm 0.7)\,10^{-8}$ | $(2.6 \pm 0.8)\,10^{-8}$ | | $(1.5 \pm 0.3)\,10^{-9}$ |
| G14.33-0.65 | $(4 \pm 4)\,10^{-10}$ | $(1.7 \pm 0.5)\,10^{-9}$ | | $(1.9 \pm 1.2)\,10^{-11}$ |
| G29.96-0.02 | $(3.3 \pm 1.4)\,10^{-10}$ | $(1.2 \pm 0.6)\,10^{-9}$ | | $(4.6 \pm 1.2)\,10^{-11}$ |
| G35.20-0.74 | $(1.2 \pm 0.5)\,10^{-9}$ | $(8 \pm 3)\,10^{-10}$ | | |
| W51 | $(1.1 \pm 0.3)\,10^{-7}$ | $(7 \pm 2)\,10^{-8}$ | | $(2.9 \pm 0.8)\,10^{-9}$ |
| ON1 | $(1.1 \pm 0.8)\,10^{-10}$ | $(2.9 \pm 1.1)\,10^{-10}$ | | |

## 6. Discussion

In this Section we discuss the main physical and chemical implications of the results presented in Sect. 5. The discussion mainly focuses on MF, DE, and EC, since F presents poor statistics, having been detected in only five sources (see Table 3) with a limited number of transitions (see Tables D.1-D.3).

### 6.1. Dust absorption effect on molecular column densities

Figure 1 shows the total column density of DE (Table 7) as a function of the observed waveband, for sources in which the molecule was detected in more than one band. Equivalent plots for MF, F, and EC are shown in Fig. E.1 of Appendix E. A clear trend of the derived column densities with the wavelength is observed: All molecules show $N_3 > N_2 > N_1$ in all sources, with a considerable gap between 3 mm and 0.9 mm values (up to ∼2 orders of magnitude). Table E.1 reports the column density ratios $N_3/N_2$ and $N_2/N_1$ in the three sources where the molecules were detected in all the three wavebands. It is $N_2/N_1 > N_3/N_2$ in all cases, by factors of ∼3 − 5 on average for all molecules (see Appendix E.2). These significant discrepancies between $N_1$, $N_2$, and $N_3$ cannot be due to differences in excitation temperature or source size, since the fits were performed with $T_{ex} = T_2$ and $\theta = \theta(2\text{ mm})$ for all wavebands (see Sects. 4 and 5.3). We interpret these results as an effect of dust opacity ($\tau_d$, see e.g. Ossenkopf & Henning 1994; Chandler & Sargent 1997; Draine 2011; Palau et al. 2014; Rivilla et al. 2017a; De Simone et al. 2020), which causes an attenuation of the molecular emission (resulting in a lower measured line intensity) of $e^{-\tau_d}$. The dust opacity depends on frequency according to $\tau_d \propto \nu^\beta$, where $\beta$ is the opacity spectral index of the source. This leads to a total column density underestimation, which becomes more and more important as the frequency increases (for instance, in our case, going from 3 mm to 0.9 mm), so that $N_2/N_1 > N_3/N_2$.

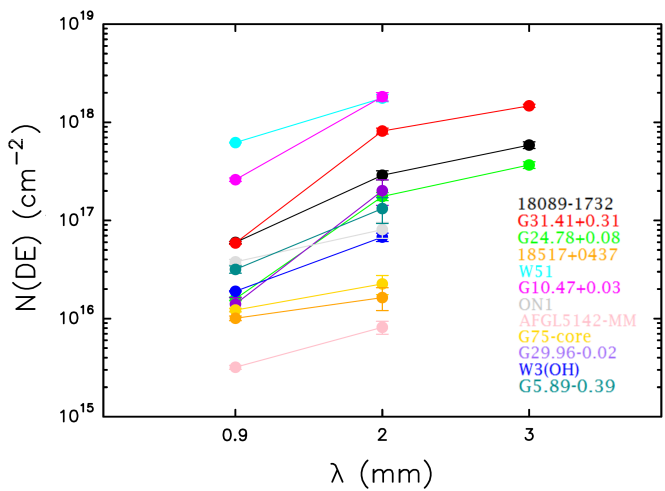

**Fig. 1.** Total molecular column densities of DE as a function of the observed waveband, in sources where the molecule was detected in more than one band.

These results highlight that the effect of dust absorption cannot be neglected when studying young and dust-rich regions such as massive star-forming cradles, in particular when comparing observations at different wavelengths. This, together with the considerations made in Sects. 4 and 5.3, brought us to concentrate our analysis on the 2 mm data. It has to be noted nonetheless that these data are still affected by dust opacity. An estimation of this attenuation is made in Appendix E.2, where a more in-depth and quantitative analysis of the dust effect on column densities, especially on their ratios in the Table E.1 sources, is performed.

### 6.2. Correlations between the molecules

In this Section we compare the derived physical parameters of the different molecules and discuss possible correlations.

#### 6.2.1. Molecular abundances

Investigating relations between molecular abundances might give us important clues about the formation processes of COMs (see e.g. Yamamoto 2017). In Fig. 2 we compare the abundances relative to $H_2$ of MF, DE, and EC (Table 10), derived from the respective column densities at 2 mm (see Sect. 5.5). For each pair of tracers, we have performed a linear regression fit to the data to check a possible correlation between the different abundances. A very strong correlation emerges between each pair of molecules (linear correlation coefficient $r \geq 0.92$), spanning 2-3 orders of magnitude in abundance (∼$10^{-10} - 10^{-7}$ for MF and DE, ∼$10^{-11} - 10^{-9}$ for EC), which are uniformly covered by our source sample. We also compare our results with measurements obtained in different interstellar environments, including other high-mass star-forming regions (HMSFRs), intermediate- and low-mass star-forming regions (IMSFRs and hot corinos, respectively), a protostellar shock region (PS shock), pre-stellar cores (PCs), and Galactic centre (GC) clouds. Individual sources and respective references can be found in Table F.1. These sources agree with the correlations found in our sample, regardless of their nature. HMSFRs, in particular, show the highest abundances for all tracers, thus expanding the correlation range by ∼2 orders of magnitude. HMSFR Sgr B2(N) N2 (Belloche et al. 2016; Bonfand et al. 2017, 2019) does not appear in the plots including EC abundances (middle and lower panels of Fig. 2, see also Sect. 6.3), since for this molecule its data points differ considerably from the distribution of all the other sources (see Sect. 6.2.2) and thus they fall out of the range shown. MF and DE (Fig. 2, upper panel) present the strongest correlation ($r = 0.99$) and rather similar abundances (i.e. a nearly constant ratio) in almost all sources, denoted by the fact that the linear best-fit to the data and the $x = y$ line nearly coincide. A strong abundance correlation between MF and DE is also found by Bisschop et al. (2007) in seven high-mass YSOs ($r = 0.90$), Brouillet et al. (2013) in Orion-KL, Jaber et al. (2014) in various objects (including hot corinos, clouds, comets) (0.95), and El-Abd et al. (2019) in the massive star-forming region NGC6334I. This result may suggest the existence of a tight physical or chemical link between these two molecules, which we further explore in Sects. 6.2.2 and 6.2.3, and thoroughly discuss in Sect. 6.4. The strong correlation we find between DE and EC (0.95, Fig. 2, lower panel) is even stronger than the one derived by Fontani et al. (2007) in six HMCs (∼0.86). It disagrees instead with Bisschop et al. (2007), who find uncorrelated abundances between DE and N-bearing species (including EC). We also find strong correlations comparing the abundances of MF, DE, and EC to F ($r > 0.9$ in all cases), although these relations are less reliable due to the poor statistics (only three sources within ∼1-2 orders of magnitudine in molecular abundance). For MF and F, this would agree with what found by Jaber et al. (2014) ($r = 0.92$).

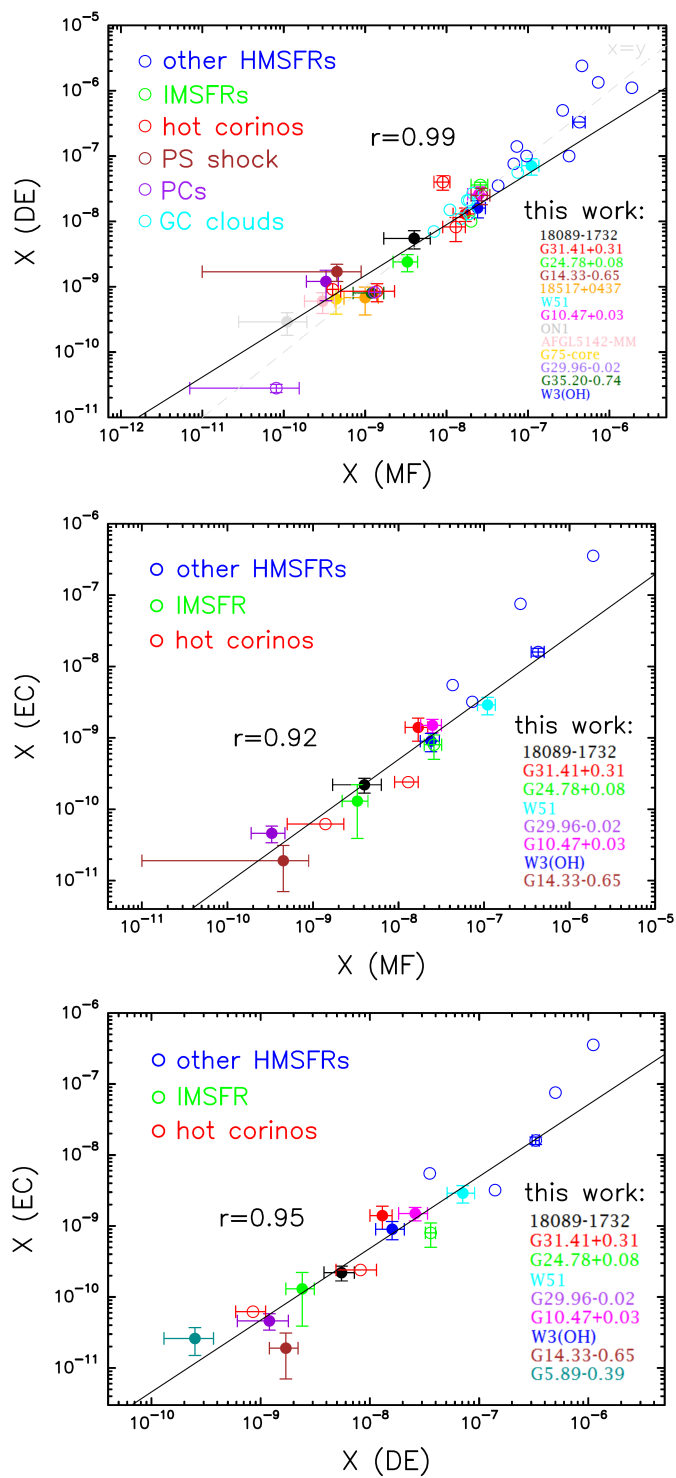

**Fig. 2.** Comparison between the observed molecular abundances (*X*, Table 10) of MF and DE (upper panel), MF and EC (middle panel), and DE and EC (lower panel). The sources analysed in this work are drawn with filled coloured circles, while literature ones (see Sect. 6.2.1 and Table F.1) with empty coloured circles. All abundances are relative to $H_2$. Error bars are shown whenever available. The solid black line corresponds to the linear best-fit to the data of the sources studied in this work, while the dashed grey line to the identity. The linear correlation coefficient between the two molecules (*r*) is also given.

### 6.2.2. Molecular ratios

Molecular ratios are considered one of the main tools to investigate potential chemical links between COMs (see e.g. Bisschop et al. 2007; Fontani et al. 2007; Rivilla et al. 2017a; El-Abd et al. 2019). Table 11 shows the 2 mm column density ratios MF/DE, MF/EC, and DE/EC, in all sources for which at least two of the species have been detected. In Fig. 3 we report the molecular ratios derived from our analysis together with literature values from other types of sources (see Table F.1) as a function of source luminosity. The MF/DE ratio (Fig. 3, upper panel) is remarkably constant within the errors, with values within ∼1 order of magnitude (0.2 − 3) across almost 9 orders of magnitude in luminosity ($\sim 10^{-2} - 10^7$ $L_\odot$) with a rather uniform coverage, from hot corinos to high-mass sources. Figure 4 shows the average MF/DE ratio for each type of source, extending the analysis to a protostellar shock, GC clouds, and comets (respectively from Lefloch et al. 2017, Requena-Torres et al. 2006, and Biver & Bockelée-Morvan 2019, see Table F.1 for details). All sources are consistent with a constant ratio of ∼1, even though PCs and comets report slightly higher average values (∼2). A nearly constant MF/DE ratio of ∼1 is also found by Rivilla et al. (2017a), but for only six sources (hot corinos, IMSFR and HMCs) in separate limited ranges of luminosity ($\sim 10 - 10^2$ and $\sim 10^5 - 10^6$ $L_\odot$), and by Ospina-Zamudio et al. (2018) in seven low- to high-mass sources. The MF/EC ratio (Fig. 3, middle panel) is nearly constant (∼20 on average) for the high-mass sources (black and blue circles), with values within ∼1 order of magnitude (∼4 − 40). Hot corinos show instead a higher dispersion (a factor of ∼50) between ∼2 and ∼$10^2$ $L_\odot$. Lastly, in the DE/EC ratio (Fig. 3, lower panel) high-mass sources show a slightly greater dispersion (∼2 − 92, averaging ∼30) than hot corinos (∼4 − 50). In the bottom two panels of Fig. 3, the blue data points clearly deviating from the trend of the other high-mass sources (molecular ratios < 0.5) belong to Sgr B2(N) N2, as already noted in Sect. 6.2.1.

**Table 11.** Relative column densities of MF, DE, and EC, using the column densities derived at 2 mm, $N_2$ (Tables 6, 7, and 9, respectively). Average values and standard deviation considering all sources are also given.

| Source | MF/DE | MF/EC | DE/EC |
|---|---|---|---|
| AFGL5142-MM | 0.5 ± 0.2 | | |
| 18517+0437 | 2.7 ± 1.4 | | |
| 18089-1732 | 0.9 ± 0.4 | 29 ± 12 | 33 ± 5 |
| G24.78+0.08 | 1.0 ± 0.2 | 8 ± 5 | 9 ± 5 |
| G31.41+0.31 | 2.3 ± 0.4 | 5.6 ± 1.3 | 2.5 ± 0.5 |
| G75-core | 0.5 ± 0.1 | | |
| W3(OH) | 1.5 ± 0.2 | 24 ± 3 | 16 ± 3 |
| G5.89-0.39 | | | 17 ± 9 |
| G10.47+0.03 | 1.0 ± 0.2 | 38 ± 4 | 40 ± 5 |
| G14.33-0.65 | 0.3 ± 0.2 | 24 ± 28 | 92 ± 47 |
| G29.96-0.02 | 0.2 ± 0.1 | 6.3 ± 1.8 | 34 ± 12 |
| G35.20-0.74 | 1.4 ± 0.7 | | |
| W51 | 1.5 ± 0.2 | 37 ± 4 | 24 ± 4 |
| ON1 | 0.4 ± 0.3 | | |
| average | 1.1 ± 0.7 | 21 ± 12 | 30 ± 25 |

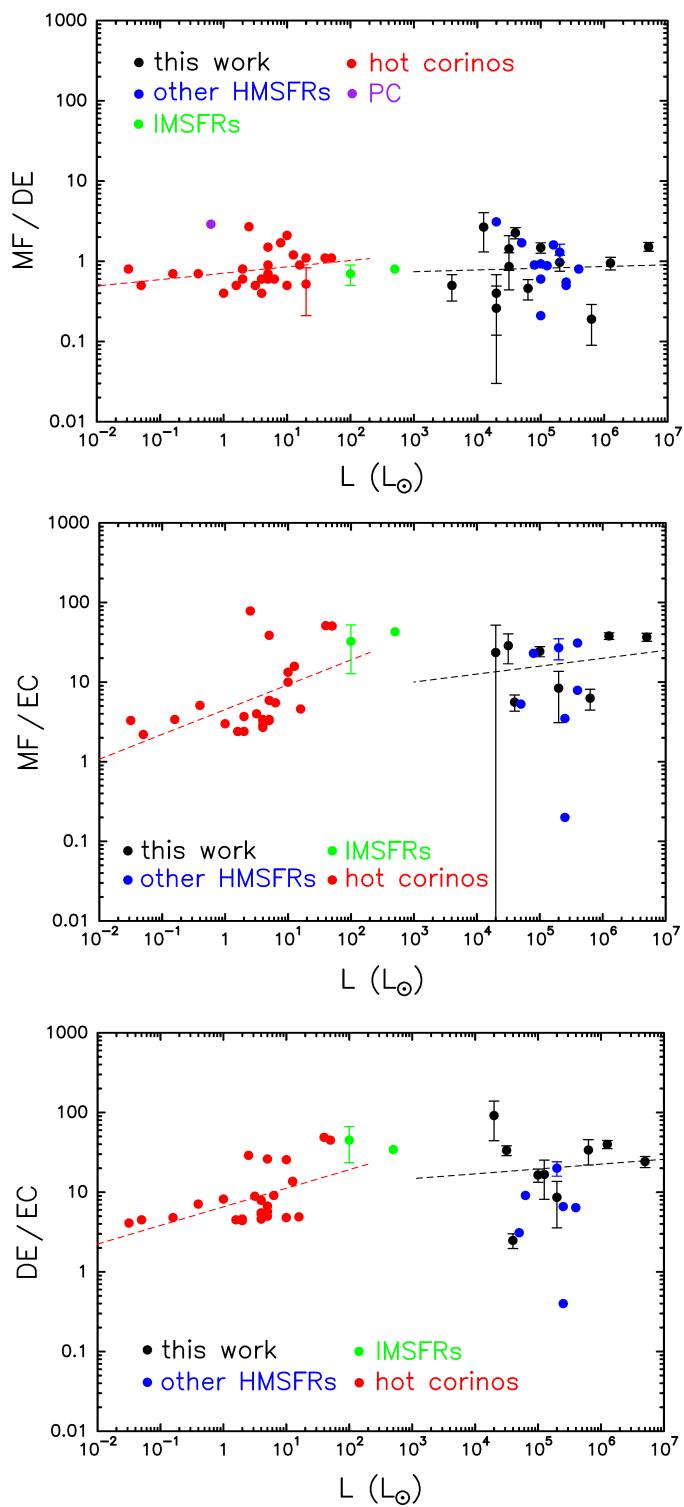

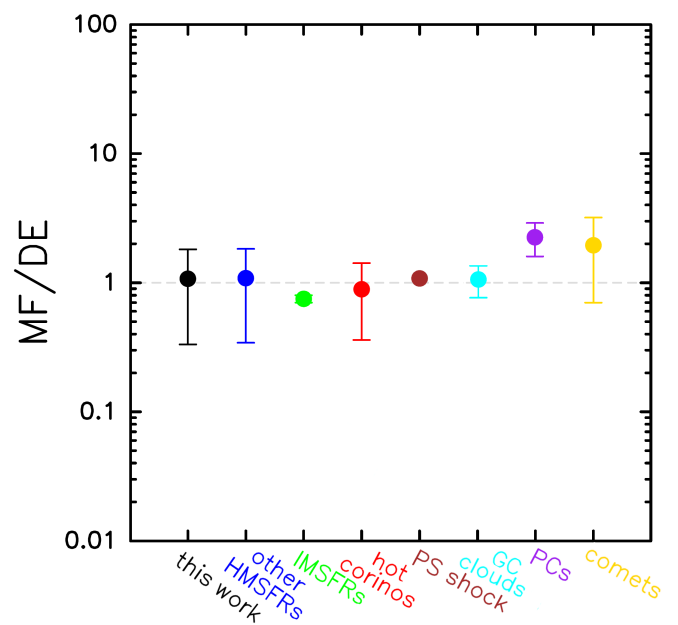

**Fig. 4.** Average MF/DE ratio compared among the sources of our sample (black) and different interstellar environments from literature (various colours, see Table F.1). Standard deviations are shown when available. The dashed grey line marks MF/DE = 1.

### 6.2.3. $\theta$, FWHM, and $T_{ex}$

In order to explore further similarities and correlations between the different molecules, we also compared other physical parameters derived in the sample from the line fitting procedure of Sect. 4 at 2 mm, such as molecular source size ($\theta$, Table 4), FWHM of the lines (Table 5), and excitation temperature ($T_2$, Table C.1).

In agreement with what already found with abundances (see Sect. 6.2.1), MF and DE show the strongest correlation ($r = 0.92$) also in terms of the estimated angular size of the emission, as depicted in Fig. 5 (upper panel), and overall cover the same range of $\theta$ ($0.8'' - 2.5''$). The pairs MF-EC and DE-EC show moderate correlations instead (0.69 and 0.59, respectively). As noted in Sect. 5.2, however, all molecules share nearly the same range of source sizes, differing by a factor of 2 at most within the same source. It has to be noted nonetheless that high angular resolution observations are needed to resolve potentially different nearby emission zones within a star-forming region and infer spatial correlations between molecules (see e.g. Mookerjea et al. 2007; Allen et al. 2017; Guzmán et al. 2018; Bøgelund et al. 2019; Belloche et al. 2020).

For FWHM, we find MF and DE (shown in Fig. 5, middle panel) to be again the most correlated ($r = 0.78$), followed by MF-EC (0.75) and DE-EC (0.63). For both MF and DE, sources W3(OH) and G10.47+0.03 show the highest FWHMs ($\sim 8 - 9$ km s$^{-1}$). Overall MF, DE, and EC share almost the same range of linewidths ($\sim 2 - 10$ km s$^{-1}$). This agrees with what found by Fontani et al. (2007) for DE and EC in G31.41+0.31, G10.47+0.03, and G29.96-0.02, and by Rivilla et al. (2017a) for MF and DE in G31.41+0.31. These results suggest, especially for MF and DE, that these molecules could trace similar gas within star-forming regions across different evolutionary stages.

**Fig. 3.** Molecular ratios MF/DE (upper panel), MF/EC (middle panel), and DE/EC (lower panel) as a function of the luminosity of the sources. The results found in this work (Table 11, black circles) are compared with a sample of different star-forming regions from literature (non-black coloured circles, see Table F.1 for references). Error bars are shown when available. The dashed black and red lines are the linear best-fits to the data of the sources included in this work and the hot corinos, respectively. For MF/EC, the large error bar of the lowest luminosity source of our sample (G14.33-0.65) results from the propagation of the high uncertainties of the individual column densities.

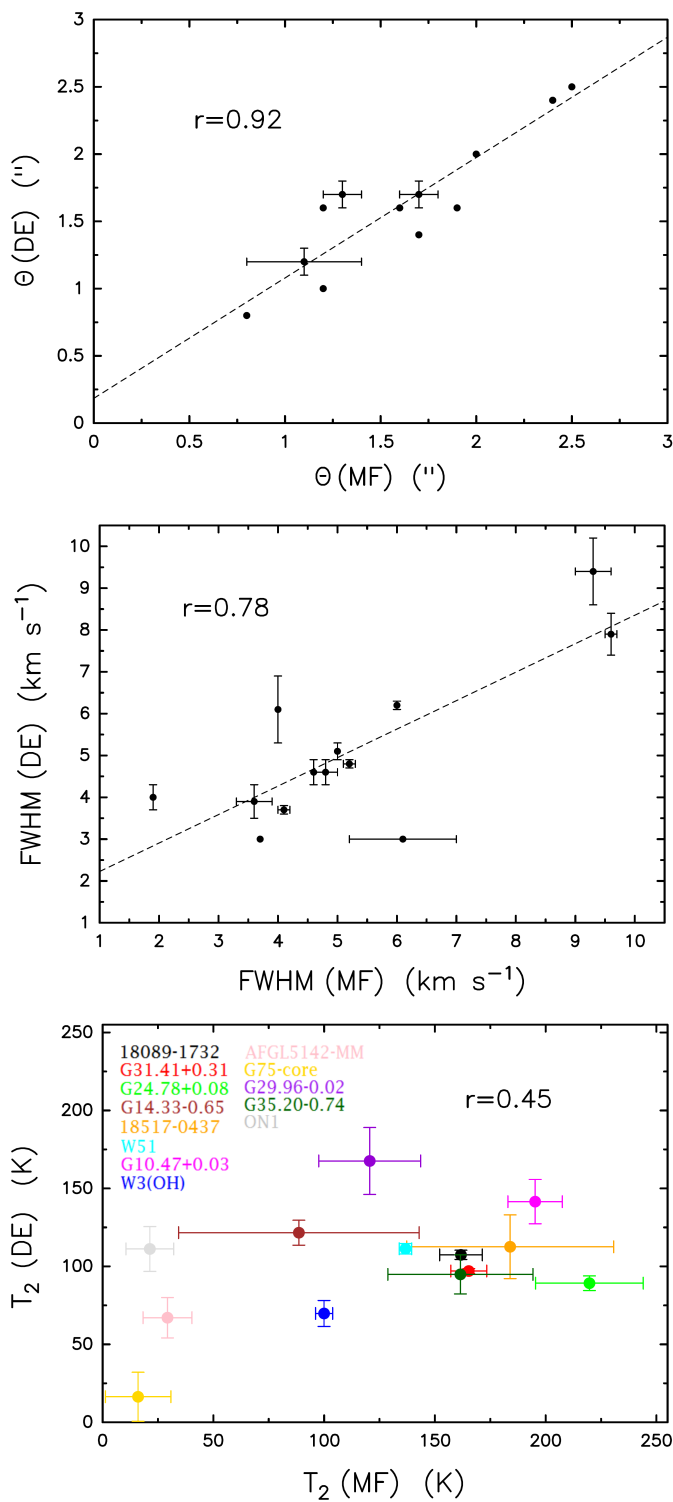

**Fig. 5.** Comparison between the source angular sizes ($\theta$, upper panel, listed in Table 4), the FWHM of the lines (middle panel, Table 5), and the excitation temperatures ($T_2$, lower panel, Table C.1) of MF and DE, obtained with the molecular line fitting procedure at 2 mm. The dashed lines are the linear best-fits to the data. The linear correlation coefficient ($r$) is also given. Values without error come from fits performed with the relative parameter fixed.

Excitation temperatures, conversely, show no significant correlations among our molecules, the only one being 0.45 between MF and DE (Fig. 5, lower panel). Moreover, temperatures and abundances of each molecule turn out to be independent, as observed by Fontani et al. (2007) for MF, DE, and EC. This attests that the strong abundance correlations found in Sect. 6.2.1 are not affected by any systematic effect due to excitation temperature. The temperature distributions of MF and EC peak at higher values ($T_2 > 150$ K) than the ones of DE and F ($T_2 < 150$ K). However, the overall temperature ranges are similar among all the molecules (especially MF, DE, and EC, see Sect. 5.3).

### 6.3. Evolution of molecular abundances

In this Section we evaluate the variation of the derived molecular abundances (Table 10) with the evolutionary stage of the sources, in order to draw an evolutionary sequence and potentially infer the most likely formation pathways for the COMs. We report detections of COMs at 2 mm in sources at three different evolutionary stages (HMPO, INT, and UCHII, see Sect. 2 and Table 1). MF and DE have been detected at all three stages, while EC only in INT and UCHII sources, and F only in INTs, so the analysis mainly focuses on the first three molecules. As it can be noted from Table 1, the three groups represent different luminosity ranges: $\sim 10^3 - 10^4\, L_\odot$ the HMPOs, $\sim 10^4 - 10^5\, L_\odot$ the INTs, and $\sim 10^4 - 10^7\, L_\odot$ the UCHIIs. We can interpret this distribution on the basis of the theoretical model developed by Molinari et al. (2008) for young massive stellar objects, predicting an increase in the total luminosity of the clump during the protostellar phase (mainly due to accretion), and a gradual stabilisation following the ignition of the (proto)star. However, since luminosity can depend not only on age but also on mass, we use the ratio L/M as an evolutionary tracer, which is expected to increase with evolution. For the sources of our sample, the mass was estimated assuming a spherical shape via the formula:

$$M = \frac{4}{3}\pi \left(\frac{\overline{D}}{2}\right)^3 \frac{N(H_2)}{\overline{D}}\, m(H)\, \mu\,, \quad (2)$$

where $\overline{D}$ is the linear source diameter corresponding to the angular source size $\overline{\theta}$ (see Sect. 5.2 and Table 4), $N(H_2)$ is the molecular hydrogen column density, $m(H) = 1.7 \cdot 10^{-24}$ g is the mass of the atomic hydrogen, and $\mu = 2.8$ (Kauffmann et al. 2008) is the mean molecular weight per hydrogen molecule. For each source, $N(H_2)$ has been rescaled to the respective $\overline{\theta}$ (see Sect. 5.5). Errors on $L/M$ were computed assuming a 20% uncertainty both on luminosity and mass.

Figure 6 shows the molecular abundances of MF, DE, F, and EC as a function of $L/M$ for the sources of our sample. An increasing abundance trend (with a similar slope of the linear fit) is evident in all molecules, spanning up to $\sim 3$ orders of magnitude both in abundance and $L/M$. MF and DE abundances nearly coincide, consistently with the $\sim 1$ constant ratio found in Sect. 6.2.2. These trends are analysed in detail in Fig. 7 for individual molecules, where we introduce the evolutionary classification of the sources of our sample and compare our results with a sample of various interstellar environments (see Table F.1). We assumed as abundance uncertainty, when not available, a conservative factor 3 above and below the value, in order to cover $\sim 1$ order of magnitude in total. The full sample shows increasing abundances going from lower to higher $L/M$ (Fig. 7). This behaviour is mainly dominated by the total luminosity and is not affected by any distance-induced observational bias, since we have checked that molecular abundances are independent from source mass and distance. PC and most HMSFRs are in very good agreement with the trend observed in our sample, while hot corinos, IMSFRs and some of the other HMSFRs show slightly higher values.

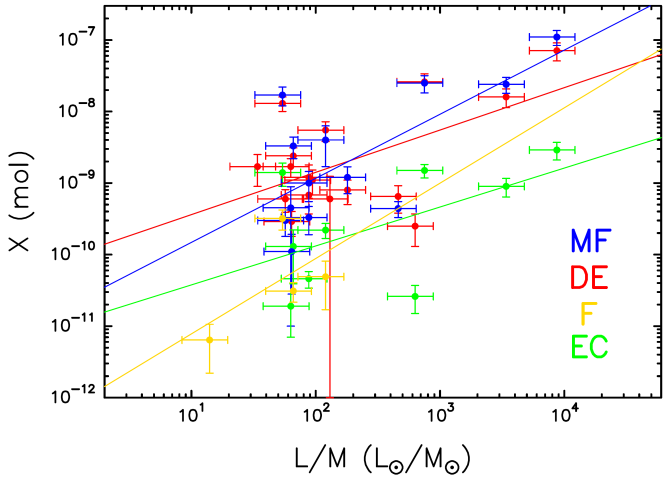

**Fig. 6.** Abundances relative to $H_2$ of MF, DE, F, and EC (Table 10) as a function of the total luminosity/mass ratio (see Sect. 6.3 for details) for the sources of our sample. The linear best-fit to the data is shown for each molecule.

Since the results of the latter are based on interferometric data, this discrepancy could be due to the different angular resolution. Although we accounted for beam dilution effects as consistently as possible (see Sect. 4), lower resolution (single-dish) observations may still result in slightly underestimated molecular column densities.

Figure 8 summarises the main result of this analysis, showing the average abundances of the four molecules with respect to the evolutionary stage of the sources. For molecules detected at multiple stages (MF, DE, and EC), average values increase with the evolution, namely from protostellar to intermediate until UCHII regions, preserving the mutual molecular ratios. The increasing trend is particularly evident for MF and DE. Average abundances increasing with time were also found by Gerner et al. (2014) for less complex molecules $CH_3OH$ (methanol), $CH_3CN$ (methyl cyanide), and other simpler molecules, and were predicted by Choudhury et al. (2015) for COMs including MF and DE through evolutionary models of HMCs.

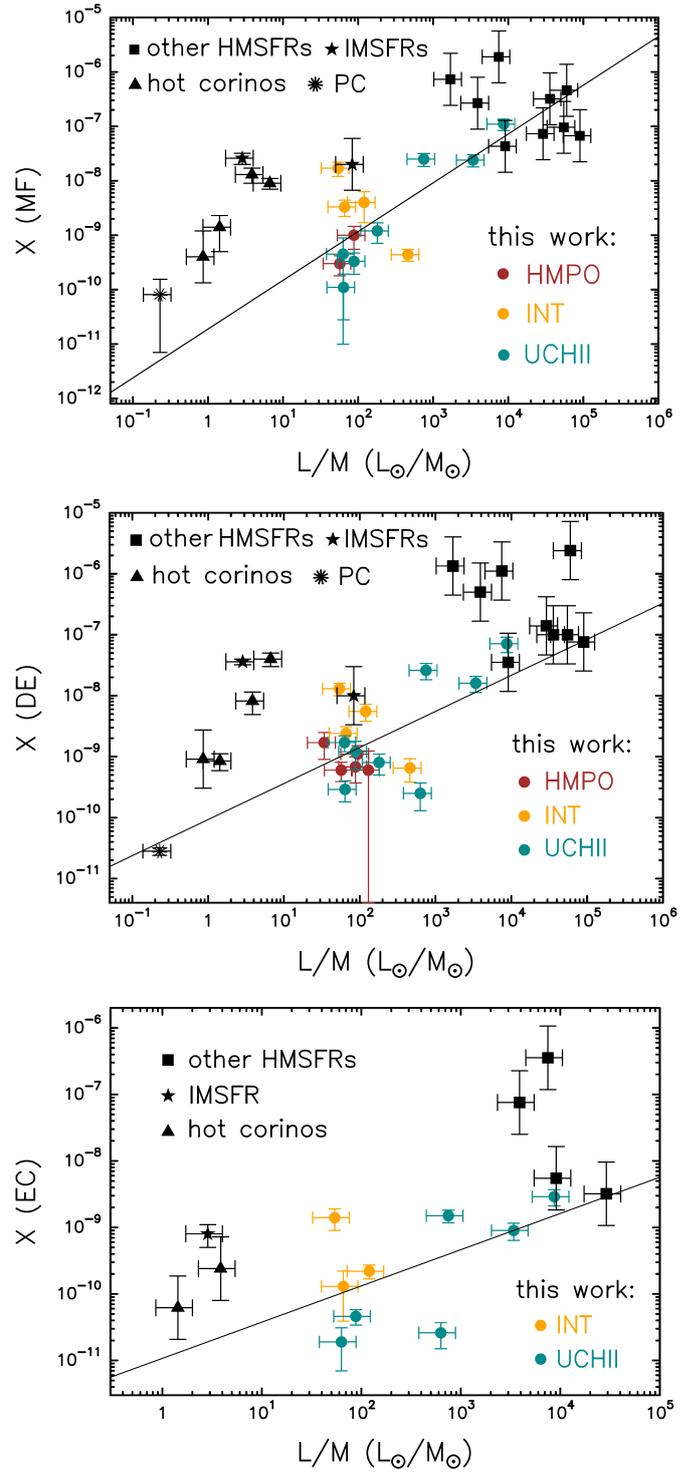

**Fig. 7.** Same as Fig. 6, but for individual molecules MF (upper panel), DE (middle panel), and EC (lower panel). The evolutionary classification is shown for the sources of our sample (different colours), while black symbols represent different interstellar sources taken from literature for comparison (see Table F.1 for references). The black lines fit the data of the sources included in this work.

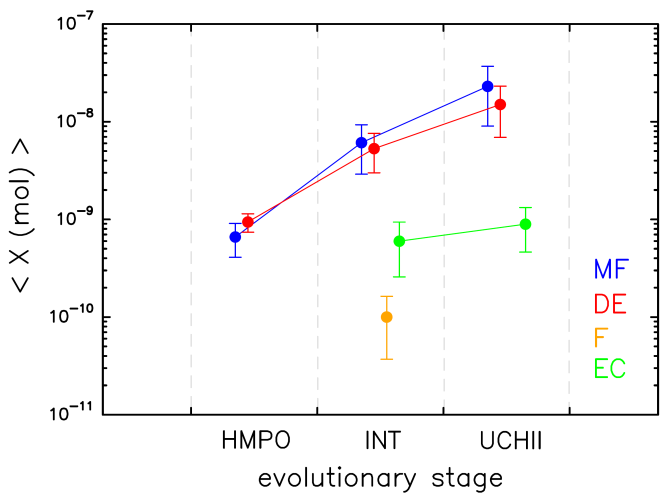

**Fig. 8.** Average abundances relative to $H_2$ (with respective standard errors) of MF, DE, F, and EC (different colours), as a function of the evolutionary stage.

### 6.4. Implications for the chemistry of COMs

The abundances of MF, DE, and EC are very well correlated ($r \geq 0.92$, Fig. 2) and their mutual molecular ratios are nearly constant (Figs. 3-4). The result is very robust since it is based on a sample with good statistics (20 sources in our sample plus 59 sources from literature overall), covering several orders of magnitude in abundance and source luminosity.

In some cases, this may indicate a chemical link between the species. This is most likely the case of MF and DE, showing the strongest correlations in many parameters (abundance, source size, and FWHM) and a constant $\sim 1$ ratio over a remarkable $\sim 9$ orders of magnitude in source luminosity (Fig. 3, upper panel), with a limited scatter both in a large sample of low- to high-mass star-forming regions and among different interstellar environments (Fig. 4). The link may consist in a common formation pathway or in one species being the precursor of the other. The first scenario is indeed predicted by the theoretical model of Garrod & Herbst (2006) and Garrod et al. (2008), who propose a common formation route through surface chemistry on dust grains at low temperatures ($\leq 50$ K), from the methoxy precursor $CH_3O$ (see also Allen & Robinson 1977):

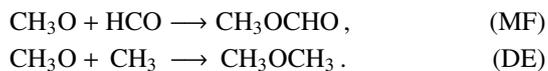

$CH_3O + HCO \longrightarrow CH_3OCHO$, (MF)
$CH_3O + CH_3 \longrightarrow CH_3OCH_3$. (DE)

Balucani et al. (2015) present instead a gas-phase route able to efficiently form MF from DE at low temperatures ($\sim 10$ K) through reactions involving the radical $CH_3OCH_2$:

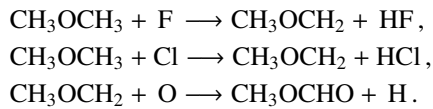

$CH_3OCH_3 + F \longrightarrow CH_3OCH_2 + HF$,
$CH_3OCH_3 + Cl \longrightarrow CH_3OCH_2 + HCl$,
$CH_3OCH_2 + O \longrightarrow CH_3OCHO + H$.

In addition, the correlated FWHM of the lines (middle panel of Fig. 5), the similar overall range of excitation temperatures (Sect. 5.3), and the spatial coexistence derived from interferometric observations (e.g. Brouillet et al. 2013; Bøgelund et al. 2019; El-Abd et al. 2019) suggest that MF and DE could trace the same gas in various environments and evolutionary stages.

However, also in the case of species for which a chemical link is not so clear (EC and MF, or EC and DE, showing slightly higher dispersion in molecular ratios, Fig. 3, bottom two panels) a clear abundance trend is observed. A potential link between these molecules may involve the methyl radical $CH_3$ as a common precursor. EC could indeed form through a sequence of gas-phase and grain-surface reactions mainly involving the CN and $CH_3$ radicals (Garrod et al. 2017). We cannot exclude either the existence of a chemical link with formamide, consistent with the abundance correlations ($> 0.9$) found in Sect. 6.2.1, but the poor statistics obtained for this molecule prevents conclusive considerations, and needs to be improved by further targeted observations. Although the formation paths of formamide are still under debate (see e.g. Bisschop et al. 2007; Barone et al. 2015; Codella et al. 2017; Skouteris et al. 2017; Ligterink et al. 2018; Quénard et al. 2018; López-Sepulcre et al. 2019), recent works propose that it would form more efficiently on icy dust grains during the cold phases of star formation (Jones et al. 2011; López-Sepulcre et al. 2015; Fedoseev et al. 2016). It has to be noted, however, that abundance correlations between molecules do not necessarily imply the existence of a chemical link, as recently proved by Quénard et al. (2018) for formamide and HNCO (isocyanic acid), and confirmed by Belloche et al. (2020) in a sample of hot corinos. These observational correlations seem to be a necessary but not sufficient condition to claim a chemical link. Nevertheless, observations are needed to test models and understand how molecules are formed. This work shows, in fact, that between molecules whose chemistry is expected to be related (such as MF and DE) the correlations are tighter. Furthermore, a clear trend of increasing molecular abundances with $L/M$ (mainly governed by $L$) emerges for all species, spanning up to $\sim 4$ orders of magnitude in abundance and $\sim 6$ in $L/M$, which implies also a trend with the evolutionary stage of the sources (Figs. 6-8).

Besides suggesting potential individual links between the COMs, these results allow us to formulate a general, most likely scenario for their formation and evolution. The fact that the molecular ratios are nearly constant across the whole star formation process and among different types of sources is particularly interesting, because the physical conditions in these environments (especially in the case of MF/DE, Fig. 4) are different: pre-stellar cores, shock-dominated regions (protostellar shock and GC clouds), thermal-dominated regions (cores in low- to high-mass star-forming regions), and comets (whose chemical composition is thought to be presolar, see e.g. Rivilla et al. 2020). This seems to reveal a rather universal chemistry for COMs, mainly developed at the cold earliest stages of star formation and then essentially preserved through the evolution, being only marginally altered by the evolving physical conditions. In more detail, molecules may be formed in pre-stellar cores, possibly in gas phase or on the surface of dust grains, from which they can desorb thanks to non-thermal mechanisms such as cosmic rays (see e.g. Shingledecker et al. 2018; Bonfand et al. 2019; Willis et al. 2020). This would explain the detection and the relative (low) abundances in the pre-stellar cores and the comets. The lack of molecular detections (at least at 2 mm) among our 11 HMSCs may be due to the fact that they are typically much more distant than the observed PCs (which can be resolved even with relatively low resolutions, see e.g. Jiménez-Serra et al. 2016), and thus more affected by beam dilution. Later on, in star-forming regions and GC molecular clouds, other mechanisms are able to massively (and more efficiently) desorb the molecules formed on grains: thermal heating and shock-induced heating. This has the effect to significantly increase the observed gas-phase molecular abundances and thus the expected number of detections. This scenario is consistent with the trend we find between abundances and $L/M$ (proxy for the evolutionary stage), as well as with the number of detections we report for each evolutionary group (Sect. 5.1). Moreover, while low luminosity sources (pre-stellar and hot corinos) are usually isolated (or at most binary) systems, high-mass star-forming regions are clustered environments. In these regions, the thermal and shock energy injected to the medium strongly increases with time due to the protostellar activity (heating and protostellar outflows). This produces more and more desorption, accordingly increasing the gas-phase abundances of COMs with evolution. Therefore, the proposed scenario supports the formation of COMs on grain surfaces, indicating that the majority of COMs observed in star-forming regions could be produced by the desorption from icy grain mantles. However, it is still possible that gas-phase formation pathways (see e.g. Balucani et al. 2015; Codella et al. 2017; Skouteris et al. 2019), though not expected to significantly affect the molecular ratios (based on our results), could contribute to the abundance of COMs in cold regions.

Moreover, our results suggest that O- and N-bearing COMs may behave similarly in star-forming regions at all stages, sharing the same physical conditions (or even direct chemical links) for their formation. This has been found also by Fontani et al. (2007) in hot cores, whereas other authors noticed differences between O- and N-bearing COMs in both the spatial distribution (e.g. Liu (2005); Csengeri et al. (2019)) and the radial velocities (Blake et al. (1987)). We also note that, given the increasing abundance trend, molecular destruction routes seem to be less efficient than formation/desorption mechanisms, especially at later evolutionary stages (i.e. higher luminosities). However, destruction routes represent a less investigated but non-negligible topic, as they can in principle affect the predicted molecular abundances (see e.g. Garrod 2013; Shingledecker et al. 2019; Ascenzi et al. 2019 and refs. therein).

Lastly, we stress that the angular resolution of our data (Table 2) is larger than the size of the observed sources. Although this issue has been taken into account through the beam dilution factor applied in the line fitting procedure (see Sect. 4), we are still not able to spatially resolve the inner structure of the targets, which is often fragmented into multiple smaller objects in potentially diverse evolutionary stages. The observed emission could hence include contributions from both the inner hot core and its cooler outer envelope, preventing a clear distinction between nearby emission zones, and causing sometimes potentially misleading correlations among differently distributed molecules. High angular resolution interferometric observations would be able to confirm more robustly the proposed scenario for the formation of COMs, as they can more accurately identify spatial correlations and resolve the potential protostellar multiplicity within a region (see e.g. Murillo et al. 2018). Nevertheless, we do not find relevant differences by comparing our results to interferometric data from literature, seemingly indicating that the observed chemistry is almost the same across different spatial scales within star-forming regions.

## 7. Summary and conclusions

In this work we have analysed spectra at 3, 2, and 0.9 mm of 39 selected high-mass star-forming regions at different evolutionary stages (HMSCs to UCHIIs) obtained with the IRAM-30m telescope, searching for rotational transitions of the complex O-bearing molecules $CH_3OCHO$ (MF) and $CH_3OCH_3$ (DE), and N-bearing molecules $NH_2CHO$ (F) and $C_2H_5CN$ (EC). We have reported molecular detections in 20 sources, performing a line fitting procedure to derive the main physical parameters for each molecule. We summarise below the main results of this study:

- The highest number of detections was reported in UCHII regions (45%, 9 out of 20 sources). DE was detected in 19 sources, while MF in 13, EC in 9, and F in 5.
- We observe relevant discrepancies between the total molecular column densities obtained at different wavelengths (up to 2 orders of magnitude between 0.9 and 3 mm), so that in all sources $N_3(3\,mm) > N_2(2\,mm) > N_1(0.9\,mm)$ and $N_2/N_1 > N_3/N_2$. This can be interpreted as an effect of the differential attenuation caused by dust opacity at each frequency ($\tau_d \propto \nu^\beta$), proving that dust properties have indeed to be considered when dealing with young, tipically dust-rich star-forming regions at multiple wavelengths. Therefore, we chose the 2 mm data for our analysis (being the band that reported the most detections) and found source-averaged column densities ranging from $\sim 10^{15}$ to $\sim 10^{18}$ cm$^{-2}$ for MF, DE, and EC, and from $\sim 10^{14}$ to $10^{17}$ cm$^{-2}$ for F.
- The derived abundances with respect to $H_2$ are $\sim 10^{-10} - 10^{-7}$ for MF and DE, $\sim 10^{-12} - 10^{-10}$ for F, and $\sim 10^{-11} - 10^{-9}$ for EC. For all species we find a consistent overall range of linewidths ($\sim 2 - 10$ km s$^{-1}$) and excitation temperatures ($\sim 20 - 220$ K).
- We find very strong correlations ($r \geq 0.92$) between the abundances of MF, DE, and EC, spanning $\sim 3$ orders of magnitude in abundance, uniformly covered by our sample. We have compared our results with heterogeneous sources from literature (including low-, intermediate- and high-mass star-forming regions, a protostellar shock region, pre-stellar cores and Galactic centre clouds), which confirmed and expanded the correlations to $\sim 4$ orders of magnitude in abundance for all tracers. We also find nearly constant molecular ratios with respect to source luminosity across all evolutionary stages and among different types of sources, indicating that the chemistry of COMs is mainly developed at early stages and then preserved during the evolution, barely altered by the changing local physical conditions. These results may suggest a potential link between MF, DE, and EC, whereas for F, though consistent with correlations ($r > 0.9$), we cannot draw conclusions due to the poor statistics. In particular, we claim that MF and DE are most likely chemically linked, since they show the strongest correlation in most parameters (abundance, FWHM, and source size) and a remarkably constant ratio of $\sim 1$ across a wide variety of sources at all evolutionary stages (also including comets), spanning a striking $\sim 9$ orders of magnitude in luminosity. The link may consist in a common formation pathway, such as from precursor $CH_3O$ as predicted by Garrod & Herbst (2006) and Garrod et al. (2008), or in one species being the precursor of the other, as proposed by Balucani et al. (2015) with MF forming from DE. MF-EC and DE-EC may share $CH_3$ as common precursor instead (see e.g. Beuther et al. 2007). Although observational correlations alone are not enough to prove a chemical link, this work shows that they are tighter among molecules whose chemistry is expected to be related (e.g. MF and DE).
- We have also evaluated the variation of molecular abundances with the evolutionary stage of the source (traced by the luminosity/mass ratio) finding a clear increasing trend for all species over up to 6 orders of magnitude in $L/M$, ranging from pre-stellar cores and hot corinos to UCHIIs.
- Based on correlations, molecular ratios and evolutionary trend, we propose a general scenario for the formation and evolution of COMs, which involves a prevalent formation at low temperatures in the earliest phases of star formation (likely mainly on frozen dust grains) followed by a growing desorption powered by the progressive thermal and shock-induced heating of the core with evolution. This would explain the increasing observed gas-phase abundances and number of molecular detections. Moreover, these results suggest that O- and N-bearing COMs might have a similar behaviour in star-forming regions at all stages. Interestingly, this analysis also points out that molecular abundances might serve as evolutionary tracers within the whole star formation process.

In conclusion, we stress that the physical parameters derived in our sample represent average values across the whole clumps, and could therefore include also contributions from outside the cores. Relevant improvements to this work will come from high angular resolution observations, able to resolve the inner structure of these regions and hence to better locate the molecular emission, allowing to more accurately identify spatial correlations between COMs. In particular, interferometric observations of a large sample of star-forming regions in different evolutionary stages, like the one studied in this work, will be able to confirm and improve the proposed scenario for the formation and evolution of COMs.

*Acknowledgements.* We thank the IRAM-30m staff for the precious help during the different observing runs. V.M.R. has received funding from the European Union's Horizon 2020 research and innovation programme under the Marie Skłodowska-Curie grant agreement No 664931. L.C. acknowledges support from the Italian Ministero dell'Istruzione, Università e Ricerca through the grant Progetti Premiali 2012 - iALMA (CUP C52I13000140001).

## Appendix A: Sources without detections

In this Appendix we list the 19 sources of the initial sample of 39 (see Sect. 2) which did not report detections of the COMs analysed in this work (MF, DE, F, and EC).

**Table A.1.** List of the observed sources of the original sample without detections of any of the COMs studied in this work (see Sect. 2).

| Source | $\alpha$(J2000) (h : m : s) | $\delta$(J2000) (° : ′ : ″) | $d$ (kpc) | Classification | References |
|---|---|---|---|---|---|
| I00117-MM1 | 00 : 14 : 26.1 | +64 : 28 : 44 | 1.8 | HMPO | (1, 2, 3, 4, 5, 6, 7, 8) |
| I00117-MM2 | 00 : 14 : 26.3 | +64 : 28 : 28 | 1.8 | HMSC | (1, 2, 3, 4, 5, 6, 7) |
| AFGL5142-EC | 05 : 30 : 48.7 | +33 : 47 : 53 | 1.8 | HMSC | (1, 3, 4, 5, 6, 7, 9, 15) |
| 05358-mm1 | 05 : 39 : 13.1 | +35 : 45 : 51 | 1.8 | HMPO | (1, 3, 4, 5, 6, 7) |
| 18264-1152M1 | 18 : 29 : 14.6 | −11 : 50 : 22 | 3.5 | HMPO | (8, 10, 11) |
| G028-C3(MM11) | 18 : 42 : 44.0 | −04 : 01 : 54 | 5.0 | HMSC | (3, 7) |
| G028-C1(MM9) | 18 : 42 : 46.9 | −04 : 04 : 08 | 5.0 | HMSC | (1, 2, 3, 7) |
| G034-F2(MM7) | 18 : 53 : 16.5 | +01 : 26 : 10 | 3.7 | HMSC | (1, 2, 3, 7) |
| G034-F1(MM8) | 18 : 53 : 19.1 | +01 : 26 : 53 | 3.7 | HMSC | (1, 2, 3, 7) |
| G034-G2(MM2) | 18 : 56 : 50.0 | +01 : 23 : 08 | 2.9 | HMSC | (1, 2, 3, 7) |
| I19035-VLA1 | 19 : 06 : 01.5 | +06 : 46 : 35 | 2.2 | UCHII | (1, 2, 3, 4, 5, 6, 7) |
| I20293-WC | 20 : 31 : 10.7 | +40 : 03 : 28 | 2.0 | HMSC | (1, 2, 3, 4, 5, 6, 7) |
| I21307 | 21 : 32 : 30.6 | +51 : 02 : 16 | 3.2 | HMPO | (1, 2, 3, 4, 5, 6, 7) |
| I22134-B | 22 : 15 : 05.8 | +58 : 48 : 59 | 2.6 | HMSC | (1, 2, 3, 4, 5, 6, 7) |
| I22134-VLA1 | 22 : 15 : 09.2 | +58 : 49 : 08 | 2.6 | UCHII | (1, 2, 3, 4, 5, 6, 7) |
| I22134-G | 22 : 15 : 10.5 | +58 : 48 : 59 | 2.6 | HMSC | (1, 2, 3, 4, 5, 6, 7) |
| 22198+6336 | 22 : 21 : 26.8 | +63 : 51 : 37 | 0.7 | HMPO | (8, 12, 13, 14) |
| 23033+5951 | 23 : 05 : 24.6 | +60 : 08 : 09 | 3.5 | UCHII | (1, 2, 3, 4, 5, 6, 7, 8) |
| NGC7538-IRS9 | 23 : 14 : 01.8 | +61 : 27 : 20 | 2.8 | UCHII | (1, 2, 3, 4, 5, 6, 7, 8, 15) |

**References.** [1] Fontani et al. 2011; [2] Fontani et al. 2014; [3] Fontani et al. 2015a; [4] Fontani et al. 2015b; [5] Fontani et al. 2016; [6] Fontani et al. 2018; [7] Colzi et al. 2018a; [8] Colzi et al. 2018b; [9] Mininni et al. 2018; [10] Fazal et al. 2008; [11] Leurini et al. 2007; [12] Jin et al. 2016; [13] Sánchez-Monge et al. 2010; [14] Fujisawa et al. 2014; [15] Fontani et al. 2019.

## Appendix B: Selected fits

In this Appendix we show, for each molecule, selected transitions detected with the molecular line fitting procedure (see Sect. 4) performed in different wavebands and sources.

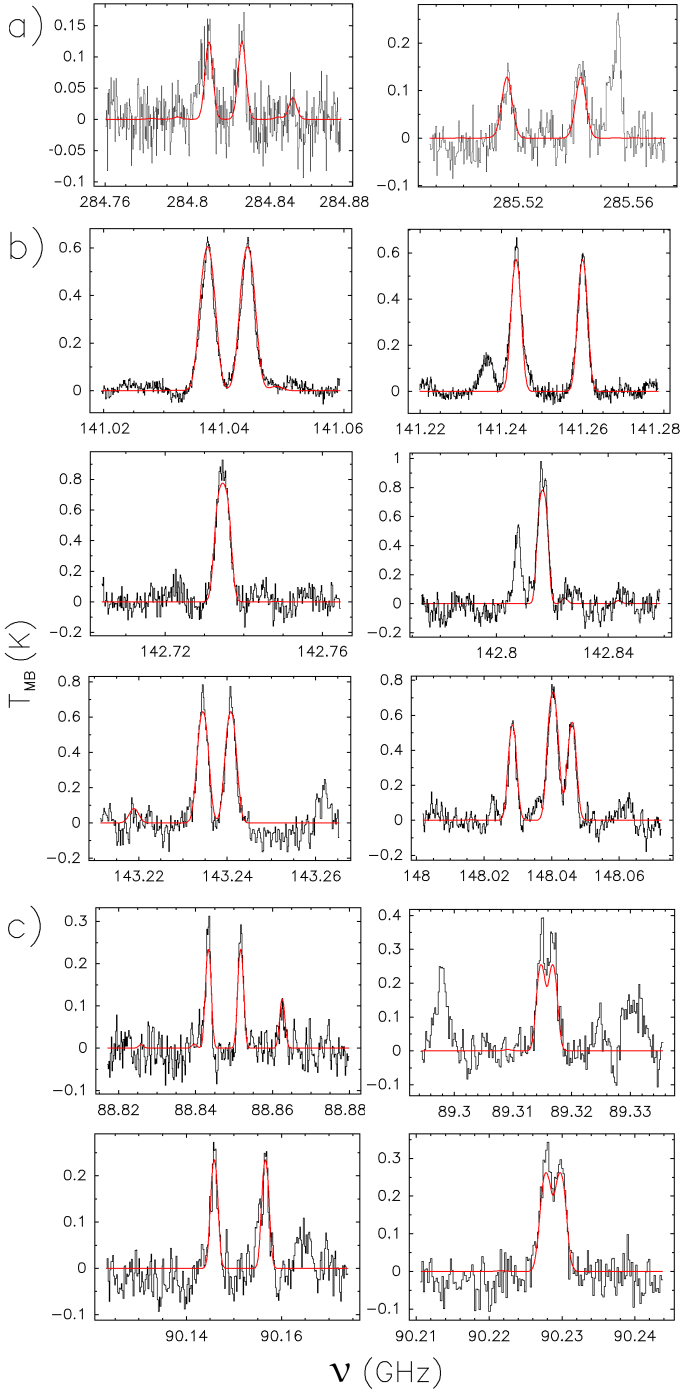

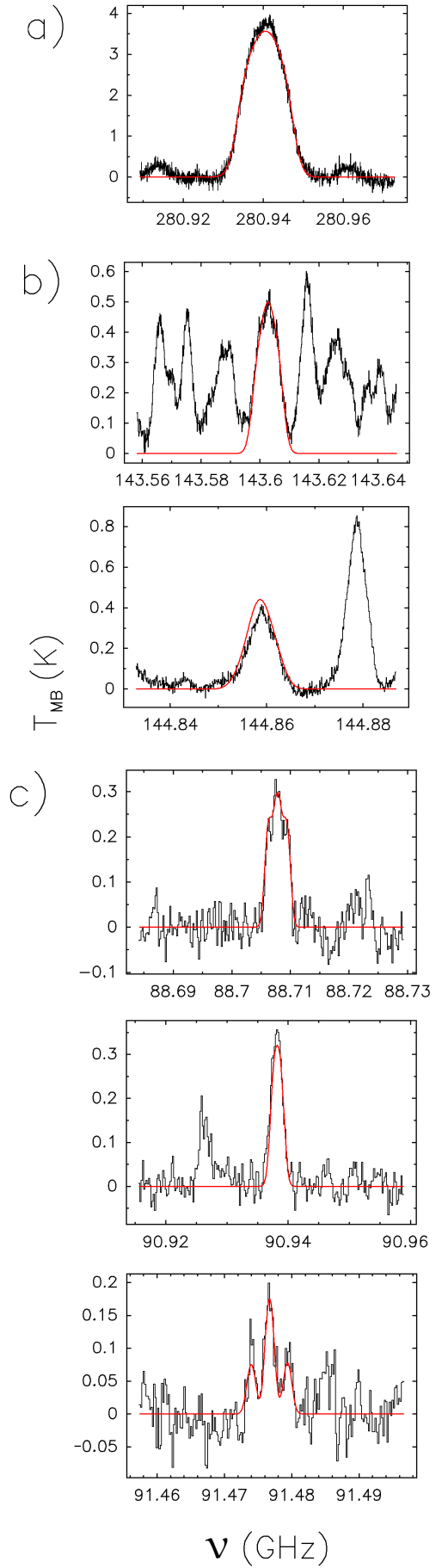

**Fig. B.1.** Selected transitions of MF detected in different wavebands and sources: **a)** 0.9 mm waveband, source 18089-1732; **b)** 2 mm, G31.41+0.31; **c)** 3 mm, G31.41+0.31. The LTE synthetic spectrum obtained in the line fitting procedure with MADCUBA (see Sect. 4) is overplotted in red. See Tables D.1-D.3 for a list of the brightest lines detected for each molecule in each waveband and their spectroscopic parameters.

**Fig. B.2.** Same as Fig. B.1, but for DE: **a)** 0.9 mm waveband, source W51; **b)** 2 mm, G10.47+0.03; **c)** 3 mm, G31.41+0.31.

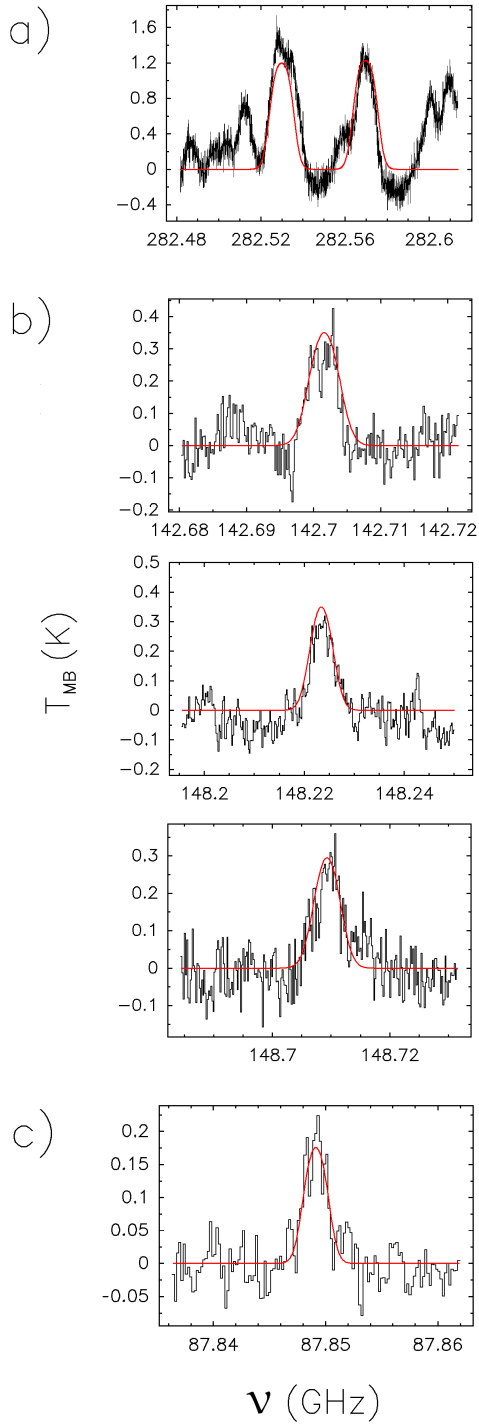

**Fig. B.3.** Same as Figs. B.1-B.2, but for F: **a)** 0.9 mm waveband, source W51; **b)** 2 mm, G31.41+0.31; **c)** 3 mm, G31.41+0.31.

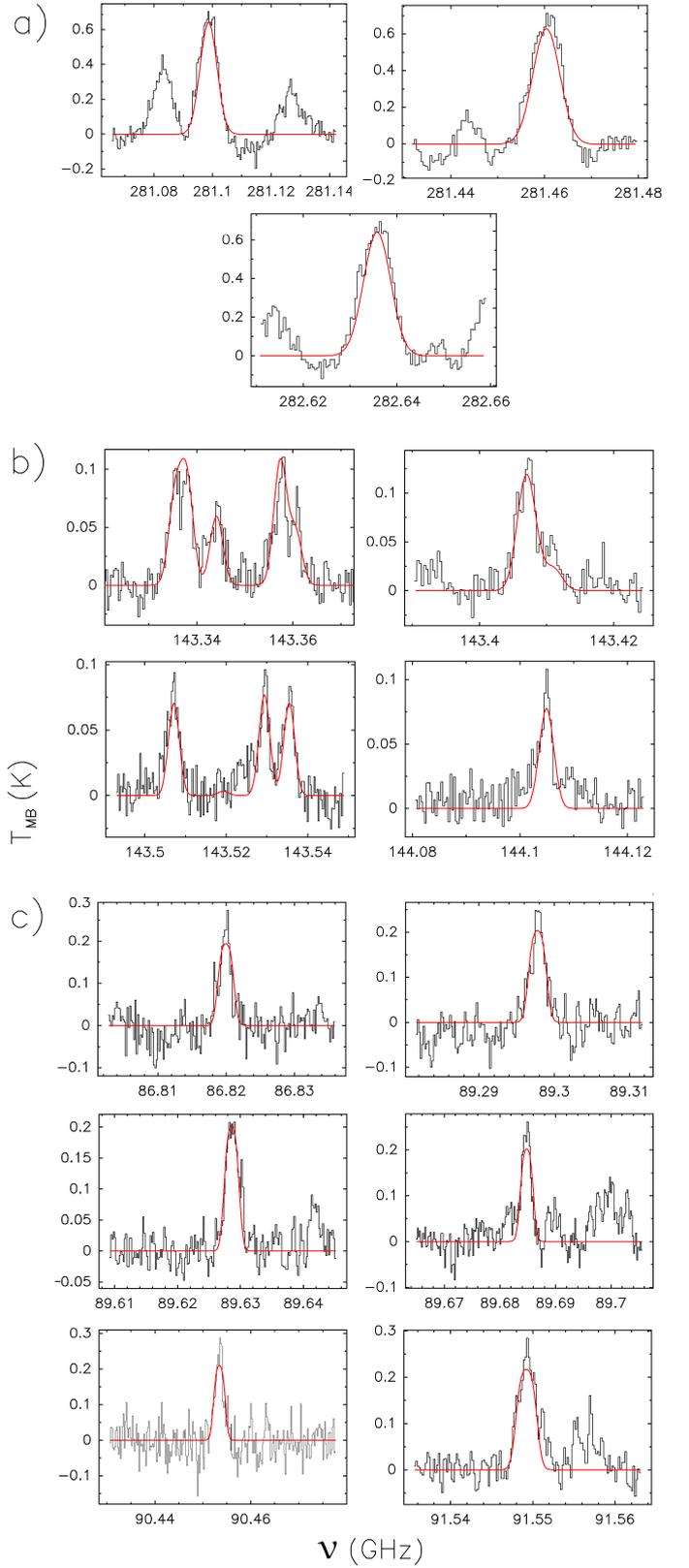

**Fig. B.4.** Same as Figs. B.1-B.3, but for EC: **a)** 0.9 mm waveband, source G10.47+0.03; **b)** 2 mm, G29.96-0.02; **c)** 3 mm, G31.41+0.31.

# Appendix C: Other physical parameters obtained from the fits

In this Appendix we report the results for the physical parameters derived from the molecular line fitting procedure (see Sect. 4) not included in Sect. 5.

*Appendix C.1: Excitation temperatures*

Table C.1 shows the excitation temperatures ($T_{ex}$, Sect. 5.3) obtained for each molecule in the different wavebands assuming LTE conditions.

**Table C.1.** Excitation temperatures of MF, DE, F, and EC, obtained from the fits (see Sects. 4 and 5.3) in the three observed wavebands: $T_1$ (0.9 mm), $T_2$ (2 mm), and $T_3$ (3 mm). Values without error come from fits performed with the $T_{ex}$ parameter fixed. Here and in the following table, the horizontal black lines subdivide the sources according to their evolutionary classification (see Table 1).

| Source | $T_{ex}$(MF) (K) | | | $T_{ex}$(DE) (K) | | | $T_{ex}$(F) (K) | | | $T_{ex}$(EC) (K) | | |
|---|---|---|---|---|---|---|---|---|---|---|---|---|
| | $T_1$ | $T_2$ | $T_3$ | $T_1$ | $T_2$ | $T_3$ | $T_1$ | $T_2$ | $T_3$ | $T_1$ | $T_2$ | $T_3$ |
| 05358-mm3 | | | | 105 ± 42 | | | | | | | | |
| AFGL5142-MM | | 29 ± 11 | | 99 ± 30 | 67 ± 13 | | | | | | | |
| 18182-1433M1 | | | | | 88 ± 13 | | | | | | | |
| 18517+0437 | 85 ± 21 | 184 ± 47 | | 116 ± 16 | 113 ± 20 | | | | | | | |
| I20293-MM1 | | | | | 134 ± 19 | | | | | | | |
| I23385 | | | | | 83 ± 40 | | | | | | | |
| 18089-1732 | 191 ± 8 | 162 ± 10 | 128 ± 38 | 81 ± 11 | 107 ± 3 | 67 ± 4 | | 100 ± 41 | | 73 | 179 ± 10 | 87 ± 30 |
| G24.78+0.08 | 114 ± 44 | 220 ± 24 | 173 ± 49 | 110 ± 14 | 89 ± 5 | 75 ± 10 | 97 | | 82 ± 27 | 87 | 72 ± 13 | 94 ± 29 |
| G31.41+0.31 | 177 ± 12 | 165 ± 8 | 155 ± 11 | 106 ± 4 | 97 ± 2 | 84 ± 5 | 93 | 115 ± 19 | 143 ± 33 | 124 ± 75 | 203 ± 16 | 223 ± 38 |
| G75-core | | 28 ± 5 | | 63 ± 29 | 34 ± 15 | | | | | | | |
| 20126+4104M1 | | | | | | | | 88 ± 34 | | | | |
| W3(OH) | 188 ± 47 | 100 ± 4 | | 107 ± 7 | 70 ± 8 | | | | | 175 | 193 ± 19 | |
| G5.89-0.39 | | | | | 87 ± 14 | | | | | | 29 ± 4 | |
| G10.47+0.03 | 84 ± 11 | 195 ± 12 | | 159 ± 9 | 142 ± 14 | | | | | 65 ± 19 | 137 ± 6 | |
| G14.33-0.65 | | 89 ± 54 | 100 ± 53 | | 122 ± 8 | | | | | | 81 ± 42 | |
| G29.96-0.02 | | 121 ± 23 | | 111 ± 50 | 168 ± 21 | | | | | | 131 ± 13 | |
| G35.20-0.74 | | 162 ± 33 | | | 95 ± 13 | | | | | | | |
| W51 | 133 ± 5 | 137 ± 4 | | 120 ± 2 | 111 ± 3 | | 88 ± 2 | | | 83 ± 44 | 162 ± 4 | |
| 19410+2336 | | | | 150 ± 34 | | | | | | | | |
| ON1 | | 21 ± 11 | | 46 ± 40 | 111 ± 14 | | | | | | | |

*Appendix C.2: Systemic velocities*

Table C.2 reports the best-fit LSR source velocities ($V_{LSR}$) obtained for each molecule in the 2 mm waveband.

**Table C.2.** The LSR source velocities ($V_{LSR}$) obtained for each molecule in the 2 mm waveband. Values without error come from fits performed with the $V_{LSR}$ parameter fixed.

| Source | $V_{LSR}$ (km s$^{-1}$) | | | |
|---|---|---|---|---|
| | MF | DE | F | EC |
| AFGL5142-MM | −2.5 ± 0.1 | −2.3 ± 0.1 | | |
| 18182-1433M1 | | 59.1 | | |
| 18517+0437 | 44.1 ± 0.4 | 44.0 | | |
| I20293-MM1 | | 6.3 ± 0.3 | | |
| I23385 | | −49.8 | | |
| 18089-1732 | 32.6 ± 0.1 | 32.7 ± 0.1 | 32.4 ± 0.3 | 33.7 ± 0.1 |
| G24.78+0.08 | 110.8 | 111.1 ± 0.1 | 111.4 ± 0.2 | 110.2 |
| G31.41+0.31 | 97.3 ± 0.1 | 97.5 ± 0.1 | 97.4 ± 0.1 | 97.2 ± 0.1 |
| 20126+4104M1 | | | −4.0 | |
| G75-core | −0.2 ± 0.1 | −0.5 ± 0.2 | | |
| W3(OH) | −47.9 ± 0.1 | −46.8 ± 0.2 | | −47.6 ± 0.1 |
| G5.89-0.39 | | 9.2 ± 0.2 | | 9.8 ± 0.4 |
| G10.47+0.03 | 66.0 ± 0.1 | 67.2 ± 0.3 | | 66.8 ± 0.1 |
| G14.33-0.65 | 22.6 | 22.9 ± 0.2 | | 22.5 |
| G29.96-0.02 | 97.7 | 97.6 ± 0.2 | | 97.6 ± 0.1 |
| G35.20-0.74 | 32.3 | 32.2 | | |
| W51 | 55.9 ± 0.1 | 56.4 ± 0.1 | | 57.9 ± 0.1 |
| ON1 | 13.0 | 11.9 ± 0.3 | | |

# Appendix D: Detected molecular transitions

In this Appendix we list the most intense rotational transitions (considering all the sources) detected with MADCUBA (see Sect. 4) for each molecule in each waveband.

**Table D.1.** Selection of the transitions which were detected for each molecule at 2 mm and used for the fits. The spectral parameters are taken from the JPL catalogue for MF lines and the CDMS catalogue for DE, F, and EC lines. We show transitions with $T_{MB} > 0.1$ K (for MF and EC) and $T_{MB} > 0.3$ K (for DE); every detected transition of F is present instead.

| Frequency (GHz) | Transition | $\log A_{ij}^{(a)}$ (s$^{-1}$) | $E_{up}^{(b)}$ (K) | Frequency (GHz) | Transition | $\log A_{ij}^{(a)}$ (s$^{-1}$) | $E_{up}^{(b)}$ (K) |
|---|---|---|---|---|---|---|---|
| | MF | | | | DE | | |
| 141.037702 | 12(2, 11) − 11(2, 10) | −4.396 | 47 | 141.832255 | 8(3, 5) − 8(2, 6) | −4.853 | 45 |
| 141.044354 | 12(2, 11) − 11(2, 10) | −4.396 | 47 | 141.835507 | 8(3, 5) − 8(2, 6) | −4.853 | 45 |
| 141.244026 | 11(3, 8) − 10(3, 7) | −4.412 | 46 | 143.020781 | 3(2, 2) − 2(1, 1) | −4.862 | 11 |
| 141.260421 | 11(3, 8) − 10(3, 7) | −4.412 | 46 | 143.163002 | 13(2, 12) − 13(1, 13) | −5.010 | 88 |
| 141.652995 | 11(2, 9) − 10(2, 8) | −4.392 | 43 | 143.599415 | 7(3, 4) − 7(2, 5) | −4.856 | 38 |
| 141.667012 | 11(2, 9) − 10(2, 8) | −4.392 | 43 | 143.602992 | 7(3, 4) − 7(2, 5) | −4.856 | 38 |
| 142.733524 | 13(1, 13) − 12(1, 12) | −4.368 | 49 | 143.606236 | 7(3, 4) − 7(2, 5) | −4.856 | 38 |
| 142.735139 | 13(1, 13) − 12(1, 12) | −4.368 | 49 | 144.858991 | 6(3, 3) − 6(2, 4) | −4.872 | 32 |
| 142.815476 | 13(0, 13) − 12(0, 12) | −4.368 | 49 | 144.862041 | 6(3, 3) − 6(2, 4) | −4.868 | 32 |
| 142.817021 | 13(0, 13) − 12(0, 12) | −4.367 | 49 | 145.547165 | 16(1, 15) − 16(0, 16) | −5.024 | 127 |
| 143.234201 | 12(1, 11) − 11(1, 10) | −4.374 | 47 | 145.680397 | 5(3, 2) − 5(2, 3) | −4.922 | 26 |
| 143.240505 | 12(1, 11) − 11(1, 10) | −4.374 | 47 | 145.682677 | 5(3, 2) − 5(2, 3) | −4.896 | 26 |
| 146.977678 | 12(3, 10) − 11(3, 9) | −4.356 | 52 | 146.166246 | 4(3, 1) − 4(2, 2) | −5.100 | 22 |
| 146.988047 | 12(3, 10) − 11(3, 9) | −4.356 | 52 | 146.677951 | 4(3, 2) − 4(2, 3) | −5.096 | 22 |
| 148.028088 | 12(6, 6) − 11(6, 5) | −4.442 | 70 | 146.704743 | 3(2, 1) − 2(1, 2) | −4.856 | 11 |
| 148.039433 | 12(6, 7) − 11(6, 6) | −4.441 | 70 | 146.872547 | 5(3, 3) − 5(2, 4) | −4.914 | 26 |
| 148.040699 | 12(6, 7) − 11(6, 6) | −4.441 | 70 | 147.024902 | 7(1, 7) − 6(0, 6) | −4.719 | 26 |
| 148.045822 | 12(6, 6) − 11(6, 5) | −4.441 | 70 | 147.025599 | 7(1, 7) − 6(0, 6) | −4.719 | 26 |
| 148.516039 | 12(5, 8) − 11(5, 7) | −4.395 | 63 | 147.206816 | 6(3, 4) − 6(2, 5) | −4.855 | 32 |
| 148.545009 | 12(5, 8) − 11(5, 7) | −4.412 | 63 | 147.21074 | 6(3, 4) − 6(2, 5) | −4.852 | 32 |
| 148.614838 | 12(5, 7) − 11(5, 6) | −4.411 | 63 | 147.731365 | 7(3, 5) − 7(2, 6) | −4.828 | 38 |
| 148.664523 | 12(5, 7) − 11(5, 6) | −4.394 | 63 | 147.734969 | 7(3, 5) − 7(2, 6) | −4.827 | 38 |
| 148.79779 | 12(4, 9) − 11(4, 8) | −4.362 | 57 | 148.497096 | 8(3, 6) − 8(2, 7) | −4.807 | 45 |
| 148.805941 | 12(4, 9) − 11(4, 8) | −4.361 | 57 | 148.500397 | 8(3, 6) − 8(2, 7) | −4.807 | 45 |
| 151.950079 | 13(2, 12) − 12(2, 11) | −4.297 | 55 | 148.503843 | 8(3, 6) − 8(2, 7) | −4.807 | 45 |
| 151.956625 | 13(2, 12) − 12(2, 11) | −4.296 | 55 | | EC | | |
| 153.350475 | 14(1, 14) − 13(1, 13) | −4.273 | 57 | 142.34633 | 16(2, 15) − 15(2, 14) | −3.624 | 63 |
| 153.352035 | 14(1, 14) − 13(1, 13) | −4.273 | 57 | 143.335284 | 16(8, 8) − 15(8, 7) | −3.733 | 130 |
| 153.397844 | 14(0, 14) − 13(0, 13) | −4.273 | 57 | 143.335284 | 16(8, 9) − 15(8, 8) | −3.733 | 130 |
| 153.399352 | 14(0, 14) − 13(0, 13) | −4.273 | 57 | 143.33771 | 16(7, 10) − 15(7, 9) | −3.701 | 113 |
| 153.512752 | 13(1, 12) − 12(1, 11) | −4.282 | 55 | 143.33771 | 16(7, 9) − 15(7, 8) | −3.701 | 113 |
| 153.518739 | 13(1, 12) − 12(1, 11) | −4.282 | 55 | 143.343925 | 16(9, 7) − 15(9, 6) | −3.774 | 148 |
| 153.553231 | 12(2, 10) − 11(2, 9) | −4.284 | 51 | 143.343925 | 16(9, 8) − 15(9, 7) | −3.774 | 148 |
| 153.56692 | 12(2, 10) − 11(2, 9) | −4.284 | 51 | 143.357203 | 16(6, 11) − 15(6, 10) | −3.674 | 98 |
| | F | | | 143.357203 | 16(6, 10) − 15(6, 9) | −3.674 | 98 |
| 140.587141 | 12(1, 11) − 12(0, 12) | −5.162 | 85 | 143.360378 | 16(10, 6) − 15(10, 5) | −3.823 | 170 |
| 142.701325 | 7(1, 7) − 6(1, 6) | −3.694 | 30 | 143.360378 | 16(10, 7) − 15(10, 6) | −3.823 | 170 |
| 146.871475 | 7(0, 7) − 6(0, 6) | −3.649 | 28 | 143.382952 | 16(11, 5) − 15(11, 4) | −3.886 | 193 |
| 148.223143 | 7(2, 6) − 6(2, 5) | −3.673 | 40 | 143.382952 | 16(11, 6) − 15(11, 5) | −3.886 | 193 |
| 148.555852 | 7(6, 2) − 6(6, 1) | −4.209 | 136 | 143.406553 | 16(5, 12) − 15(5, 11) | −3.652 | 86 |
| 148.555852 | 7(6, 1) − 6(6, 0) | −4.209 | 136 | 143.407188 | 16(5, 11) − 15(5, 10) | −3.652 | 86 |
| 148.566822 | 7(5, 3) − 6(5, 2) | −3.943 | 103 | 143.410796 | 16(12, 4) − 15(12, 3) | −3.967 | 218 |
| 148.566823 | 7(5, 2) − 6(5, 1) | −3.943 | 103 | 143.410796 | 16(12, 5) − 15(12, 4) | −3.967 | 218 |
| 148.596177 | 9(0, 9) − 8(1, 8) | −5.114 | 45 | 143.443012 | 16(13, 3) − 15(13, 2) | −4.076 | 246 |
| 148.59897 | 7(4, 4) − 6(4, 3) | −3.804 | 76 | 143.443012 | 16(13, 4) − 15(13, 3) | −4.076 | 246 |
| 148.599354 | 7(4, 3) − 6(4, 2) | −3.804 | 76 | 143.50697 | 16(4, 13) − 15(4, 12) | −3.635 | 76 |
| 148.667301 | 7(3, 5) − 6(3, 4) | −3.720 | 55 | 143.5292 | 16(3, 14) − 15(3, 13) | −3.622 | 69 |
| 148.709018 | 7(3, 4) − 6(3, 3) | −3.720 | 55 | 143.53529 | 16(4, 12) − 15(4, 11) | −3.635 | 76 |
| 153.432176 | 7(1, 6) − 6(1, 5) | −3.600 | 32 | 144.10474 | 16(3, 13) − 15(3, 12) | −3.617 | 69 |
| | | | | 145.41801 | 16(1, 15) − 15(1, 14) | −3.592 | 61 |
| | | | | 146.12004 | 16(2, 14) − 15(2, 13) | −3.590 | 64 |
| | | | | 146.894524 | 17(1, 17) − 16(1, 16) | −3.578 | 65 |
| | | | | 147.756711 | 17(0, 17) − 16(0, 16) | −3.570 | 65 |

[a] Logarithmic Einstein coefficient; [b] Rotational upper level energy.

**Table D.2.** Selection of the transitions which were detected for each molecule at 0.9 mm and used for the fits. The spectral parameters are taken from the JPL catalogue for MF lines and the CDMS catalogue for DE, F, and EC lines. We show transitions with $T_{MB} > 1$ K (for MF and DE) and $T_{MB} > 0.1$ K (for F and EC).

| Frequency (GHz) | Transition | $\log A_{ij}^{(a)}$ (s$^{-1}$) | $E_{up}^{(b)}$ (K) |
|---|---|---|---|
| MF | | | |
| 284.227369 | 23(10, 13) − 22(10, 12) | −3.548 | 229 |
| 284.243124 | 23(10, 14) − 22(10, 13) | −3.548 | 229 |
| 284.243381 | 23(10, 13) − 22(10, 12) | −3.548 | 229 |
| 284.252904 | 23(10, 14) − 22(10, 13) | −3.548 | 229 |
| 284.810313 | 23(5, 19) − 22(5, 18) | −3.478 | 181 |
| 284.826396 | 23(5, 19) − 22(5, 18) | −3.478 | 181 |
| 284.92024 | 23(9, 14) − 22(9, 13) | −3.527 | 217 |
| 284.937218 | 23(9, 15) − 22(9, 14) | −3.526 | 217 |
| 284.942751 | 23(9, 14) − 22(9, 13) | −3.526 | 217 |
| 284.945147 | 23(9, 15) − 22(9, 14) | −3.526 | 217 |
| 285.515739 | 22(5, 17) − 21(5, 16) | −3.474 | 170 |
| 285.542584 | 22(5, 17) − 21(5, 16) | −3.473 | 170 |
| 285.924822 | 23(8, 16) − 22(8, 15) | −3.506 | 206 |
| 285.940794 | 23(8, 16) − 22(8, 15) | −3.522 | 206 |
| 285.973267 | 23(8, 15) − 22(8, 14) | −3.522 | 206 |
| DE | | | |
| 280.934845 | 4(4, 1) − 3(3, 0) | −3.771 | 32 |
| 280.93916 | 4(4, 0) − 3(3, 0) | −3.771 | 32 |
| 280.93924 | 4(4, 1) − 3(3, 1) | −3.771 | 32 |
| 280.942889 | 4(4, 1) − 3(3, 0) | −3.771 | 32 |
| 280.943554 | 4(4, 0) − 3(3, 1) | −3.771 | 32 |
| F | | | |
| 282.529615 | 13(2, 11) − 12(2, 10) | −2.792 | 106 |
| 282.569429 | 13(1, 12) − 12(1, 11) | −2.785 | 98 |
| 285.750632 | 13(1, 13) − 12(0, 12) | −4.166 | 92 |
| EC | | | |
| 281.098901 | 31(2, 29) − 30(2, 28) | −2.727 | 221 |
| 281.460931 | 32(1, 31) − 31(3, 29) | −2.725 | 228 |
| 282.600634 | 33(1, 33) − 32(1, 32) | −2.718 | 233 |
| 282.636554 | 33(0, 33) − 32(0, 32) | −2.718 | 233 |
| 285.473238 | 32(3, 30) − 31(3, 29) | −2.709 | 237 |

[a] Logarithmic Einstein coefficient; [b] Rotational upper level energy.

**Table D.3.** Selection of the transitions which were detected for each molecule at 3 mm and used for the fits. The spectral parameters are taken from the JPL catalogue for MF lines and the CDMS catalogue for DE, F, and EC lines. We show transitions with $T_{MB} > 0.2$ K (for MF, DE, and EC) and $T_{MB} > 0.01$ K (for F).

| Frequency (GHz) | Transition | $\log A_{ij}^{(a)}$ (s$^{-1}$) | $E_{up}^{(b)}$ (K) |
|---|---|---|---|
| MF | | | |
| 88.843187 | 7(1, 6) − 6(1, 5) | −5.008 | 18 |
| 88.851607 | 7(1, 6) − 6(1, 5) | −5.008 | 18 |
| 89.314657 | 8(1, 8) − 7(1, 7) | −4.993 | 20 |
| 89.316642 | 8(1, 8) − 7(1, 7) | −4.993 | 20 |
| 90.145723 | 7(2, 5) − 6(2, 4) | −5.011 | 20 |
| 90.156473 | 7(2, 5) − 6(2, 4) | −5.011 | 20 |
| 90.227659 | 8(0, 8) − 7(0, 7) | −4.978 | 20 |
| 90.229624 | 8(0, 8) − 7(0, 7) | −4.978 | 20 |
| DE | | | |
| 88.707704 | 15(2, 13) − 15(1, 14) | −5.287 | 117 |
| 88.709177 | 15(2, 13) − 15(1, 14) | −5.287 | 117 |
| 90.938107 | 6(0, 6) − 5(1, 5) | −5.440 | 19 |
| F | | | |
| 86.382755 | 7(1, 6) − 7(0, 7) | −5.643 | 32 |
| 87.848874 | 4(1, 3) − 3(1, 2) | −4.367 | 14 |
| EC | | | |
| 88.323735 | 10(0, 10) − 9(0, 9) | −4.248 | 23 |
| 89.29766 | 10(2, 9) − 9(2, 8) | −4.251 | 28 |
| 89.628485 | 10(3, 8) − 9(3, 7) | −4.269 | 34 |
| 89.68471 | 10(3, 7) − 9(3, 6) | −4.268 | 34 |
| 90.453349 | 10(2, 8) − 9(2, 7) | −4.234 | 28 |
| 91.549112 | 10(1, 9) − 9(1, 8) | −4.687 | 25 |

[a] Logarithmic Einstein coefficient; [b] Rotational upper level energy.

# Appendix E: Further analysis on dust opacity in our sample

In this Appendix we expand the analysis presented in Sect. 6.1 on the dust absorption effect on column densities measured in the sources of our sample.

## Appendix E.1: Column density plots of MF, F, and EC

In addition to Fig. 1 of Sect. 6.1 showing DE, Fig. E.1 shows the total molecular column densities ($N$) of MF (Table 6), F (Table 8), and EC (Table 9), obtained in the 0.9 mm, 2 mm, and 3 mm wavebands. As already underlined in Sect. 6.1, a clear trend with the wavelength emerges for all molecules (including F even with poor statistics), highlighting the differential line intensity attenuation applied by dust in the three observed wavebands.

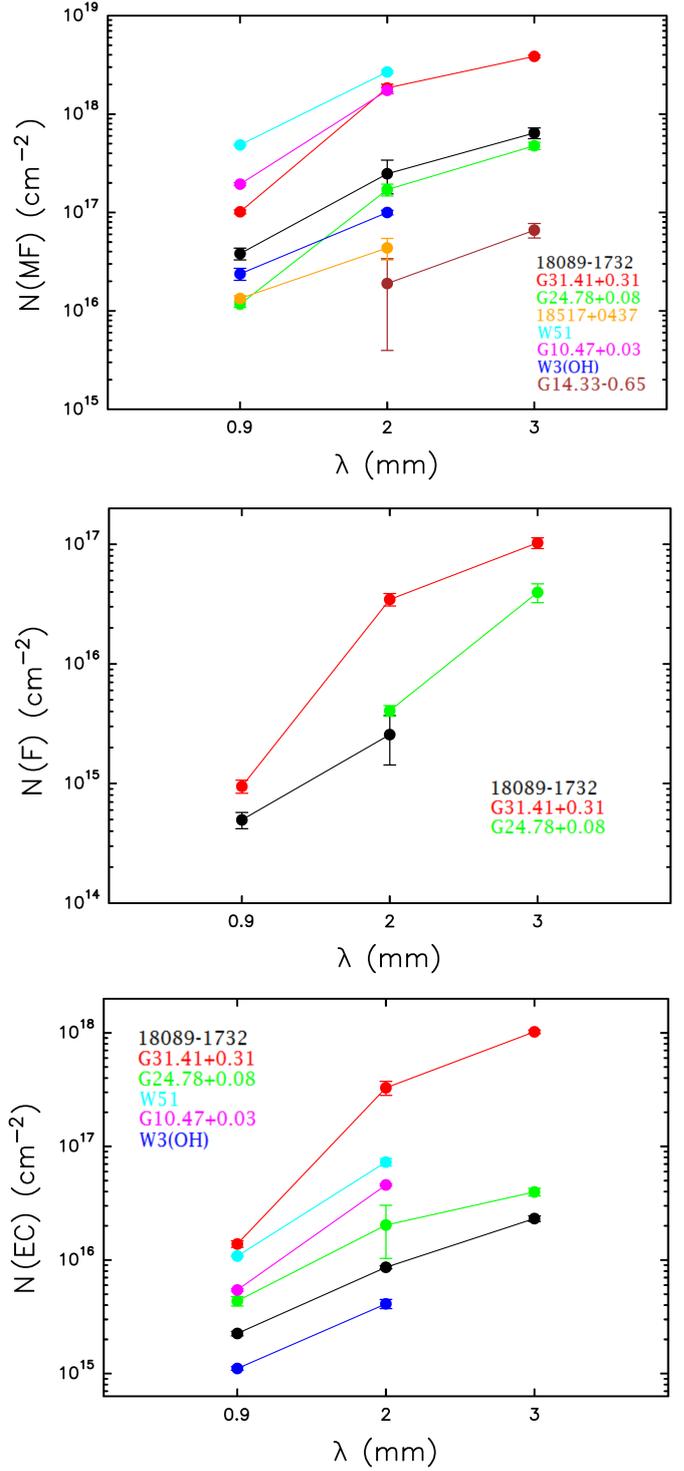

**Fig. E.1.** Total molecular column densities of MF (upper panel, listed in Table 6), F (middle panel, Table 8), and EC (lower panel, Table 9) as a function of the observed waveband, in sources where the molecule was detected in more than one band.

*Appendix E.2: Estimation of the dust opacity spectral index of 18089-1732, G24.78+0.08, G31.41+0.31 through observed column density ratios*

As anticipated in Sect. 6.1, the effect of dust absorption on molecular column density can vary from source to source, mainly due to the amount of dust. Being the dust opacity $\tau_d \propto \nu^\beta$, this in our case affects the $N_2/N_1$ ratio more than $N_3/N_2$, as it can be seen in Fig. E.2, comparing the column density ratios (listed in Table E.1) of MF, DE, F, and EC (different symbols) in 18089-1732, G24.78+0.08, and G31.41+0.31 (different colours). While $N_3/N_2$ values are very similar between the sources, a relevant differentiation emerges for $N_2/N_1$, and also a dispersion within each source between values belonging to different molecules. In theory, this latter points should coincide, depending on source opacity and excitation temperature. However, it is also possible that molecules tracing different regions within a source could face different levels of absorption based on the local conditions. From the discrepancies in the $N_2/N_1$ ratio between the three sources, it was possible to estimate their opacity spectral index ($\beta$). It can be shown, in fact, that the attenuation of the molecular line intensities caused by dust absorption is of a factor $e^{-\tau_d}$ (see e.g. Rivilla et al. 2017a), and that the ratio between the different column densities can be written as follows (see Appendix E.3):

$$\begin{cases} N_2/N_1 = exp(\tau_2[(\nu_1/\nu_2)^\beta - 1]) \\ N_3/N_2 = exp(\tau_2) \end{cases}, \quad (E.1)$$

where $\tau_2$ is the dust opacity at 2 mm, $\nu_1$ and $\nu_2$ are the central frequencies of the 0.9 and 2 mm wavebands, respectively about 285 and 145 GHz. Therefore, in Figure E.2, the horizontal blue lines identify different opacities (within a range $0.4 - 1.2$), while the slope of the dashed coloured lines is linked to the value of the spectral index $\beta$. The latter was obtained with a power regression fit to the values of each source. A higher $\beta$ (i.e. a flatter slope in this graph) implies a stronger dust absorption effect, thus a wider discrepancy between $N_2/N_1$ and $N_3/N_2$. G31.41+0.31 (red data points) shows the higher $\beta$ (2.2) and the higher $N_2/N_1$ ($\sim 14-37$), whereas G24.78+0.08 (green) gives $\beta = 1.95$, and 18089-1732 (black) $\beta = 1.55$. Table E.2 summarises the values of $\tau_2$ and $\beta$ obtained for the three sources analysed. It has to be noted that the spectral index $\beta$ depends in general on multiple factors, such as the amount of dust within the source and its properties (e.g. the grain size and shape), and the density and compactness of the source (see e.g. Miyake & Nakagawa 1993; Pollack et al. 1994; Chandler & Sargent 1997; Draine 2011). In our case, however, we do not expect relevant differences between the sources in terms of grain sizes, so the dominant parameter is the dust amount, which is proportional to the $\beta$ value estimated for each source. Based on the dust opacities at 2 mm shown in Fig. E.2 and Table E.2, we can also give an estimation of the dust absorption effect on the $N_2$ column densities we have used for the derivation of the molecular abundances, at least for the three sources included in this analysis. An attenuation of $e^{-\tau_2}$, with $\tau_2$ values between $\sim 0.6$ and $\sim 1.1$, results in a $N_2$ correction factor of $\sim 2-3$.

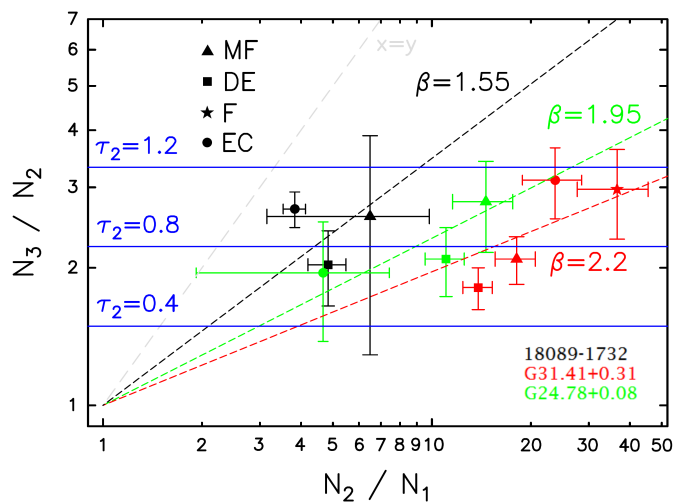

**Fig. E.2.** Comparison between the column density ratios $N_2/N_1$ and $N_3/N_2$ of MF (triangles), DE (squares), F (star), and EC (circles) in the sources 18089-1732 (black data points), G24.78+0.08 (green), and G31.41+0.31 (red). The dashed coloured lines are the power regression fits to the data of each source, and identify the related $\beta$ (see text). The horizontal blue lines correspond to different opacities at 2 mm. The dashed grey line represents the case $N_2/N_1 = N_3/N_2$ (i.e. $\beta = 1$). For further details on this plot see Sect. E.2.

**Table E.1.** Molecular column density ratios $N_3/N_2$ and $N_2/N_1$ (Tables 6-9) for the three sources where the molecules were detected in all the three wavebands, and average values.

| Source | Column density ratios | | | | | | | |
|---|---|---|---|---|---|---|---|---|
| | MF | | DE | | F | | EC | |
| | $N_3/N_2$ | $N_2/N_1$ | $N_3/N_2$ | $N_2/N_1$ | $N_3/N_2$ | $N_2/N_1$ | $N_3/N_2$ | $N_2/N_1$ |
| 18089-1732 | 2.6 ± 1.3 | 6 ± 3 | 2.0 ± 0.4 | 4.8 ± 0.6 | | 5 ± 3 | 2.7 ± 0.2 | 3.8 ± 0.3 |
| G24.78+0.08 | 2.8 ± 0.6 | 15 ± 3 | 2.1 ± 0.4 | 11.0 ± 1.5 | 10 ± 3 | | 1.9 ± 1.1 | 5 ± 3 |
| G31.41+0.31 | 2.1 ± 0.2 | 18 ± 3 | 1.8 ± 0.2 | 13.8 ± 1.4 | 3.0 ± 0.7 | 37 ± 9 | 3.1 ± 0.5 | 24 ± 5 |
| average | 2.5 ± 0.2 | 13 ± 3 | 2.0 ± 0.1 | 10 ± 2 | 6 ± 2 | 21 ± 11 | 2.6 ± 0.3 | 11 ± 5 |

**Table E.2.** Values of the dust opacity at 2 mm ($\tau_2$) and the spectral index ($\beta$) obtained for the three sources included in this analysis.

| Source | $\tau_2$ | $\beta$ |
|---|---|---|
| 18089-1732 | 0.7 – 1.0 | 1.55 |
| G24.78+0.08 | 0.7 – 1.0 | 1.95 |
| G31.41+0.31 | 0.6 – 1.1 | 2.2 |

*Appendix E.3: Derivation of Eqs. E.1 for column density ratios*

| Column density | Dust opacity |
|---|---|
| $N(3\,mm) \equiv N_3$ | $\tau(3\,mm) \equiv \tau_3$ |
| $N(2\,mm) \equiv N_2 = N_3 e^{-\tau_2}$ | $\tau(2\,mm) \equiv \tau_2 = \tau_3(\nu_2/\nu_3)^\beta$ |
| $N(0.9\,mm) \equiv N_1 = N_3 e^{-\tau_1}$ | $\tau(0.9\,mm) \equiv \tau_1 = \tau_3(\nu_1/\nu_3)^\beta$ |
| | |
| $N_2/N_1 = e^{-\tau_2}/e^{-\tau_1} = e^{-(\tau_2-\tau_1)}$ | $\tau_1 = \tau_3(\nu_2/\nu_3)^\beta(\nu_1/\nu_2)^\beta$ |
| $N_3/N_2 = N_3/(N_3 e^{-\tau_2}) = e^{\tau_2}$ | $\quad = \tau_2(\nu_1/\nu_2)^\beta$ |
| | $\tau_2 - \tau_1 = \tau_2[1 - (\nu_1/\nu_2)^\beta]$ |

$$N_2/N_1 = exp(\tau_2[(\nu_1/\nu_2)^\beta - 1])$$
$$N_3/N_2 = exp(\tau_2)$$

## Appendix F: Sources taken from the literature

In Table F.1 we report the sample of different interstellar environments taken from literature we used in the analysis of molecular correlations (Sects. 6.2.1-6.2.2) and evolutionary trend (Sect. 6.3).

**Table F.1.** List of the sources from the literature used for comparison in Sects. 6.2.1-6.2.2 and 6.3, with respective references from which the parameters used in the analysis have been taken.

| Source | $\alpha$(J2000) (h : m : s) | $\delta$(J2000) (° : ′ : ″) | References |
|---|---|---|---|
| High-mass star-forming regions (HMSFRs) ||||
| Orion KL | 05 : 35 : 14.2 | −05 : 22 : 21.5 | Taquet et al. 2015 |
| AFGL4176 | 13 : 43 : 01.7 | −62 : 08 : 51.2 | Bøgelund et al. 2019 |
| IRAS 16562-3959 CC | 16 : 59 : 41.6 | −40 : 03 : 43.6 | Guzmán et al. 2010, 2018 |
| NGC6334IRS1 | 17 : 20 : 53.0 | −35 : 47 : 02 | Bisschop et al. 2007 |
| Sgr B2(N) N3 | 17 : 47 : 19.2 | −28 : 22 : 14.9 | Bonfand et al. 2017, 2019 |
| Sgr B2(N) N4 | 17 : 47 : 19.5 | −28 : 22 : 32.4 | ″ |
| Sgr B2(N) N2 | 17 : 47 : 19.9 | −28 : 22 : 13.4 | ″ |
| Sgr B2(N) N5 | 17 : 47 : 20.0 | −28 : 22 : 41.3 | ″ |
| W33A | 18 : 14 : 38.9 | −17 : 52 : 04 | Bisschop et al. 2007 |
| G19.61-0.23 | 18 : 27 : 38.0 | −11 : 56 : 42 | Taquet et al. 2015 |
| G34.26+0.15 NE | 18 : 53 : 18.5 | +01 : 14 : 58.2 | Mookerjea et al. 2007; Rivilla et al. 2017a |
| AFGL2591 | 20 : 29 : 24.6 | +40 : 11 : 19 | Bisschop et al. 2007 |
| NGC7538IRS1 | 23 : 13 : 45.4 | +61 : 28 : 12 | ″ |
| Intermediate-mass star-forming regions (IMSFRs) ||||
| NGC7129 FIRS2 | 21 : 43 : 01.7 | +66 : 03 : 23.6 | Eiroa et al. 1998; Fuente et al. 2014; Taquet et al. 2015; Rivilla et al. 2017a |
| Cep E-A | 23 : 03 : 12.8 | +61 : 42 : 26 | Ospina-Zamudio et al. 2018 |
| Low-mass star-forming regions (Hot corinos) ||||
| L1448-2Ab | 03 : 25 : 22.4 | +30 : 45 : 13.2 | Belloche et al. 2020 |
| L1448-2A | 03 : 25 : 22.4 | +30 : 45 : 13.3 | ″ |
| L1448-NB2 | 03 : 25 : 36.3 | +30 : 45 : 15.1 | ″ |
| L1448-NB1 | 03 : 25 : 36.4 | +30 : 45 : 14.8 | ″ |
| L1448-NA | 03 : 25 : 36.5 | +30 : 45 : 21.8 | ″ |
| L1448-C | 03 : 25 : 38.9 | +30 : 44 : 05.3 | ″ |
| L1448-CS | 03 : 25 : 39.1 | +30 : 43 : 58.0 | ″ |
| IRAS2A1 | 03 : 28 : 55.6 | +31 : 14 : 37.1 | ″ |
| NGC1333 IRAS 2A | 03 : 28 : 55.6 | +31 : 14 : 37.2 | Taquet et al. 2015; Rivilla et al. 2017a |
| SVS13B | 03 : 29 : 03.1 | +31 : 15 : 51.7 | Belloche et al. 2020 |
| SVS13A | 03 : 29 : 03.8 | +31 : 16 : 03.8 | ″ |
| IRAS4A2 | 03 : 29 : 10.4 | +31 : 13 : 32.1 | ″ |
| IRAS4A1 | 03 : 29 : 10.5 | +31 : 13 : 31.0 | ″ |
| NGC1333 IRAS 4A | 03 : 29 : 10.5 | +31 : 13 : 31.1 | Taquet et al. 2015; Rivilla et al. 2017a |
| IRAS4B | 03 : 29 : 12.0 | +31 : 13 : 08.0 | Belloche et al. 2020 |
| IRAS4B2 | 03 : 29 : 12.8 | +31 : 13 : 06.8 | ″ |
| B1b-S | 03 : 33 : 21.4 | +31 : 07 : 26.4 | Gerin et al. 2015; Marcelino et al. 2018 |
| IRAM04191 | 04 : 21 : 56.9 | +15 : 29 : 46.1 | Belloche et al. 2020 |
| L1521F | 04 : 28 : 38.9 | +26 : 51 : 35.1 | ″ |
| L1527 | 04 : 39 : 53.9 | +26 : 03 : 09.7 | ″ |
| IRAS16293-2422 | 16 : 32 : 22.6 | −24 : 28 : 31.8 | Pineda et al. 2012; Jaber et al. 2014; Manigand et al. 2020 |
| SerpM-S68N | 18 : 29 : 48.1 | +01 : 16 : 43.4 | Belloche et al. 2020 |
| SerpM-S68Nb | 18 : 29 : 48.7 | +01 : 16 : 55.5 | ″ |
| SerpM-SMM4a | 18 : 29 : 56.7 | +01 : 13 : 15.6 | ″ |
| SerpS-MM18b | 18 : 30 : 03.5 | −02 : 03 : 08.3 | ″ |
| SerpS-MM18a | 18 : 30 : 04.1 | −02 : 03 : 02.5 | ″ |
| SerpS-MM22 | 18 : 30 : 12.3 | −02 : 06 : 53.6 | ″ |
| L1157 | 20 : 39 : 06.3 | +68 : 02 : 15.7 | ″ |
| GF9-2 | 20 : 51 : 29.8 | +60 : 18 : 38.4 | ″ |
| Protostellar shock region (PS shock) ||||
| L1157-B1 | 20 : 39 : 10.2 | +68 : 01 : 10 | Lefloch et al. 2017 |
| Pre-stellar cores (PCs) ||||
| B5 | 03 : 47 : 32.1 | +32 : 56 : 43.0 | Taquet et al. 2017 |
| L1544 | 05 : 04 : 17.2 | +25 : 10 : 42.8 | Doty et al. 2005; Lemke et al. 2008; Jiménez-Serra et al. 2016 |
| Galactic centre clouds (GC clouds) ||||
| MC G-0.11-0.08 | 17 : 42 : 28.0 | +29 : 02 : 55 | Requena-Torres et al. 2006 |
| MC G-0.02-0.07 | 17 : 42 : 40.0 | +28 : 58 : 00 | ″ |
| MC G+0.07-0.07 | 17 : 42 : 54.2 | +28 : 53 : 30 | ″ |
| MC G+0.24+0.01 | 17 : 42 : 59.6 | +28 : 42 : 35 | ″ |
| MC G+0.70-0.01 | 17 : 44 : 10.0 | +28 : 19 : 30 | ″ |
| MC G+0.694-0.017 | 17 : 44 : 10.0 | +28 : 20 : 05 | ″ |
| MC G+0.693-0.027 | 17 : 44 : 12.1 | +28 : 20 : 25 | ″ |
| MC G+0.62-0.10 | 17 : 44 : 18.0 | +28 : 26 : 30 | ″ |
| MC G+0.76-0.05 | 17 : 44 : 27.2 | +28 : 17 : 35 | ″ |
| MC G+0.68-0.10 | 17 : 44 : 27.2 | +28 : 23 : 20 | ″ |
| Comets ||||
| C/1995 O1 (Hale-Bopp) | | | Biver & Bockelée-Morvan 2019 |
| C/2014 Q2 (Lovejoy) | | | ″ |